\documentclass[preprint,amssymb,amsmath]{article}

\usepackage{graphicx}
\usepackage{pifont}
\usepackage{textcomp}
\graphicspath{{fig/}}
\usepackage{dcolumn}
\usepackage{bm}
\usepackage{float}
\usepackage{amssymb}
\usepackage{amsmath}
\usepackage{color}
\begin{document}

\title{Quantum Spin Pumping Mediated by Magnon}

\author{Kouki Nakata,    \\    \   \\
 Yukawa Institute for Theoretical Physics, 
Kyoto University,  \\
Kitashirakawa Oiwake-Cho, Kyoto 606-8502, Japan,  \\
nakata@yukawa.kyoto-u.ac.jp}

\maketitle

\begin{abstract}
We theoretically propose quantum spin pumping mediated by magnons,  under a time-dependent transverse magnetic field,
at the interface between a ferromagnetic insulator and a non-magnetic metal. 
The  generation of a  spin current under a thermal equilibrium condition
is discussed by calculating the spin relaxation  torque, which breaks the spin conservation law for conduction electrons
and operates the coherent magnon state.
Localized spins lose spin angular momentum by emitting  magnons
and conduction electrons flip from down to up by absorbing the momentum.
The spin relaxation torque has a resonance structure as a function of the angular frequency of the applied transverse field.
This fact is useful to enhance the spin pumping effect induced by quantum fluctuations.
We also discuss the distinction between our quantum spin pumping theory and the one proposed by Tserkovnyak et al. 
\end{abstract}

\section{Introduction}
\label{sec:intro}

A new branch of physics called spintronics
has seen a rapid development over the last decades.
The central theme is the active manipulation of spin degrees of freedom
as well as charge ones of electrons;
spintronics avoids the dissipation from Joule heating by
replacing charge currents with spin currents.
Thus establishing methods for the generation of a spin current is  significant  
from viewpoints of fundamental science and potential applications
to green information  and communication  technologies.\cite{mod}

A standard way to generate a (pure) spin current is
the spin pumping effect
at the interface between a ferromagnetic material and  a non-magnetic metal. 
There the  precession  of  the  magnetization induces a spin current pumped into a non-magnetic  metal,
which is  proportional to the rate of precession of the magnetic moment.
The precessing ferromagnet acts as a source of spin angular momentum;
spin battery.
This method has been  developed by Tserkovnyak et al.\cite{mod2} 
Now their formula   has been widely used for interpreting vast experimental results,
despite the phenomenological treatment of spin-flip scattering processes.

Thus we reformulate the spin pumping theory,
 through the Schwinger-Keldysh formalism,
to explicitly describe the spin-flip processes.
A generation of the pumped spin current is 
discussed on the basis of the spin continuity equation for conduction electrons.
Moreover  the utilization of magnons, 
which are the quantized collective  motions  of  localized spins,
has recently  been attracting considerable interest.\cite{spinwave,sandweg,Kurebayashi}
Hence, the focus of the present work lies also on  the contribution of magnons to spin pumping.
We treat localized spins as not classical variables but magnon degrees of freedom.
Therefore we can capture the (non-equilibrium) spin-flip dynamics,
where spin angular momentum is exchanged between conduction electrons and localized spins,
on the basis of  the rigorous quantum mechanical theory.
Consequently,
we can reveal the significance of time-dependent transverse magnetic fields,
which act  as quantum fluctuations,
for the spin pumping effect. 

\begin{figure}[h]
\begin{center}
\includegraphics[width=7.5cm,clip]{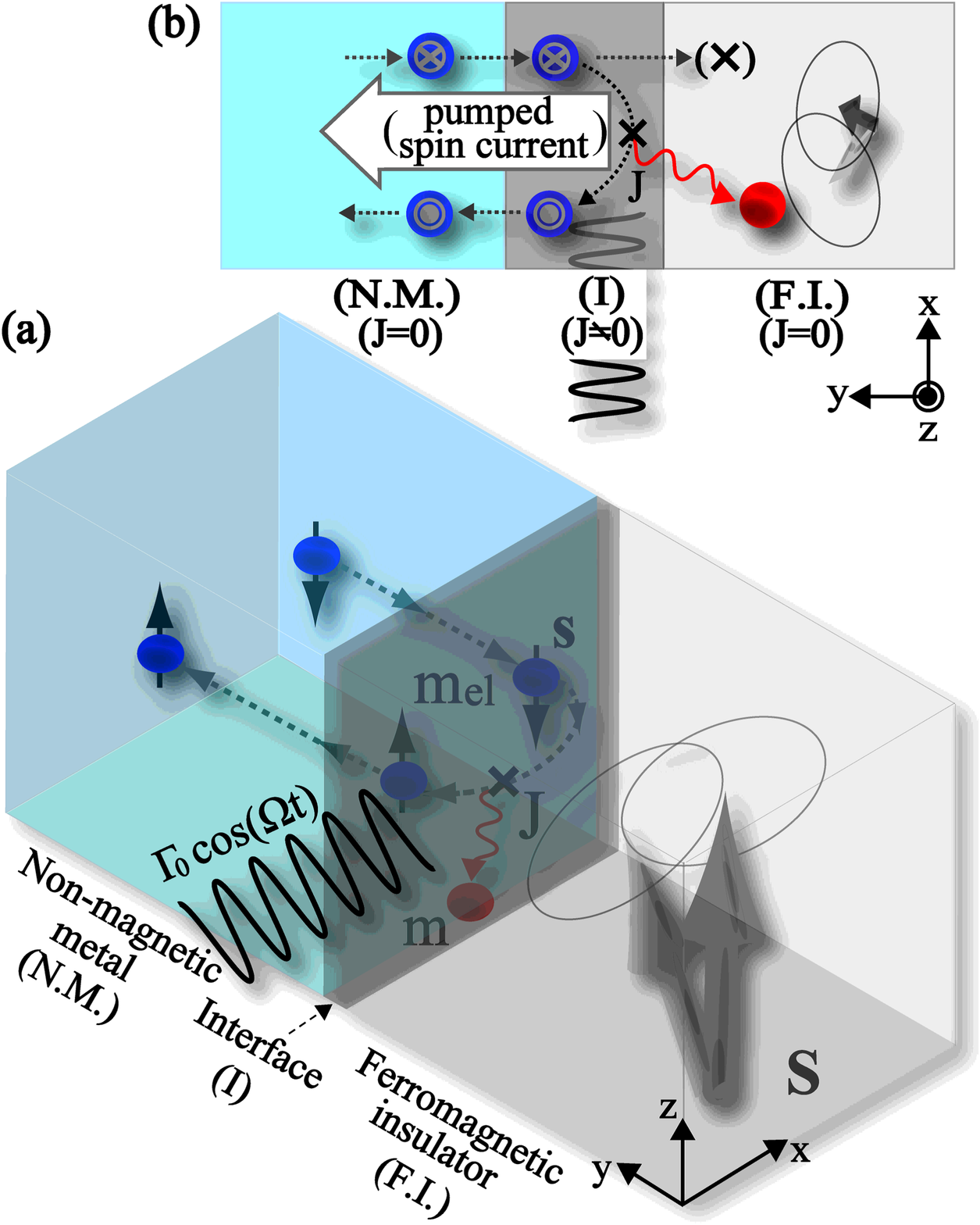}
\caption{(Color online)
 A schematic picture of quantum spin pumping mediated by magnons.
 Spheres  represent magnons and
those with arrows are conduction electrons.
(a)
Localized spins lose spin angular momentum 
by emitting  magnons
and conduction electrons flip  
from down to up by absorbing the momentum.
(b) 
An enlarged schematic    picture of the interface.
The interface is defined as an effective area where 
the Fermi gas (conduction electrons)  and 
the Bose gas (magnons) coexist to interact; $J\not=0$.
Conduction electrons  cannot  enter the ferromagnet,
which is an insulator. 
 \label{fig:pumping} }
\end{center}
\end{figure}

We consider   a ferromagnetic insulator and  non-magnetic   metal   junction\cite{spinwave}
shown in Fig. \ref{fig:pumping}
where conduction electrons couple with localized spins  $  {\mathbf{S}}({\mathbf{x}},t) $, ${\mathbf{x}}=(x, y, z) \in {\mathbf{R}}^3$, 
at the interface;
\begin{equation}
 {\cal{H}}_{\rm{ex}}=  - 2Ja_0^3  {\int_{{\mathbf{x}}\in \text{(interface)}}} d {\mathbf{x}}   {\ }  {\mathbf {S}}({\mathbf{x}},t) \cdot   {\mathbf {s}}({\mathbf{x}},t).
\label{eqn:e1}
\end{equation}
The exchange coupling constant reads  $ 2J $,  and the lattice constant of the ferromagnet is  $a_0$.
The magnitude of the interaction is supposed to be constant and 
we adopt  the continuous limit in the present study.
Conduction electron spin variables are represented as  
 \begin{equation} 
\begin{split}  
 {{s}^{j}}  &= \sum_{\eta ,\zeta  = \uparrow , \downarrow} [c^{\dagger }_{\eta} (\sigma  ^j)_{\eta \zeta} c_{\zeta}]/2    \\
                    &\equiv  (c^{\dagger }  \sigma ^j c)/2,
\end{split}  
 \end{equation}                    
where $  {  {\sigma }^j} $ are the $ 2\times 2 $  Pauli matrices;
$ [ \sigma  ^j , \sigma  ^k  ] = 2i\epsilon _{jkl} \sigma  ^l   $, ($ j, k, l = x,y,z$).
Operators $c^{\dagger }/c $ are   creation/annihilation operators for conduction electrons, 
which satisfy the (fermionic) anticommutation relation;
$ \{c_{\eta  }({\mathbf{x}}, t), c_{\zeta }^{\dagger }({\mathbf{x}}', t) \}= \delta _{\eta , \zeta } \delta ({\mathbf{x}}-{\mathbf{x'}})  $.

We focus on the dynamics at the interface where spin angular momentum is exchanged 
between conduction electrons and the ferromagnet.
We suppose the uniform magnetization and
thus  localized spin degrees of freedom   can be mapped into magnon ones 
via the Holstein-Primakoff transformation;
\begin{equation}
\begin{split}   
S^{+} ({\mathbf{x}},t)  &\equiv S^{x} ({\mathbf{x}},t)+iS^{y} ({\mathbf{x}},t)        \\
&=  \sqrt{2\tilde S}  [1- a^\dagger({\mathbf{x}},t) a({\mathbf{x}},t)/(4\tilde S) ]a({\mathbf{x}},t) + {\cal{O}}({\tilde S}^{-3/2}),
\end{split}
\end{equation}                       
   \begin{equation}
\begin{split}  
S^{-}({\mathbf{x}},t)  &\equiv S^{x} ({\mathbf{x}},t)+iS^{y} ({\mathbf{x}},t)                   \\
&=   \sqrt{2\tilde S}  a^\dagger ({\mathbf{x}},t)  [1- a^\dagger({\mathbf{x}},t) a({\mathbf{x}},t)/(4\tilde S) ]  + {\cal{O}}({\tilde S}^{-3/2}),
\end{split} 
\end{equation}            
   \begin{eqnarray}  
S^z({\mathbf{x}},t) = \tilde S-a^\dagger ({\mathbf{x}},t)  a   ({\mathbf{x}},t),        
\end{eqnarray}      
$\tilde S\equiv  S/{a_0^3}$,                             
where operators $a^{\dagger }/a $ are magnon creation/annihilation operators
satisfying the (bosonic) commutation relation; 
$ [a(\mathbf{x}, t), a^{\dagger }(\mathbf{x'}, t) ]= \delta (\mathbf{x}-\mathbf{x'}) $.
Up to the $ {\cal{O}}(S)$ terms,
localized spins  reduce to a free boson system.

Consequently in the quadratic dispersion (i.e. long wavelength) approximation, 
the  localized spin   with the applied magnetic field along the quantization axis (z-axis) is described
by the Hamiltonian $  {\cal{H}}_{\rm{mag}} $;
\begin{equation}
\begin{split}
  {\cal{H}}_{\rm{mag}}  =  \int_{{\mathbf{x}}\in \text{(interface)}} d   {\mathbf{x}}  {\ }  {a^{\dagger }(\mathbf{x},t)} 
                                             \Big(-\frac{ { \mathbf{\nabla }}^2 }{2m} + B \Big)  
                                             {a(\mathbf{x},t)},
\label{eqn:e2}
\end{split}
\end{equation}
and the Hamiltonian,  ${\cal{H}}_{\rm{ex}} (\equiv     {\cal{H}}_{\rm{ex}} ^{S}+{\cal{H}}_{\rm{ex}} ^{\prime})$, can be rewritten as
\begin{eqnarray}
 {\cal{H}}_{\rm{ex}} ^{S}   &=&   -JS  {\int_{{\mathbf{x}}\in \text{(interface)}}} d    {\mathbf{x}}  
                                                         {\ }   c^{\dagger } ({\mathbf{x}},t)  \sigma^z  c  ({\mathbf{x}},t), 
\label{eqn:e51}                                                                                                                                                               \\
{\cal{H}}_{\rm{ex}} ^{\prime} &=& -Ja_0^3  \sqrt{\frac{\tilde S}{2}} {\int_{{\mathbf{x}}\in \text{(interface)}}} d   {\mathbf{x}} 
                               \Bigg\{ a^{\dagger }({\mathbf{x}},t) \Big[1-\frac{a^{\dagger }({\mathbf{x}},t) a({\mathbf{x}},t) }{4\tilde S} \Big] 
                                       c^{\dagger }  ({\mathbf{x}},t) \sigma ^{+} c({\mathbf{x}},t )   \nonumber  \\
                                 &+ &\Big[1-\frac{a^{\dagger }({\mathbf{x}},t) a({\mathbf{x}},t) }{4\tilde S} \Big] a({\mathbf{x}},t)  
                                       c^{\dagger } ({\mathbf{x}},t)  \sigma ^{-} c({\mathbf{x}},t ) \Bigg\}.
\label{eqn:e3}
\end{eqnarray}
The variable 
 $ m $ represents the effective mass of a magnon.
We have denoted a constant applied magnetic field along the quantization axis as  $B$,
 which includes $g$-factor and Bohr magneton.
 In this paper,
 we take $\hbar =1$ for convenience. 

We have adopted  the continuous limit and hence,
as shown in Fig. \ref{fig:pumping} (b),
the interface is defined as an effective area where 
the Fermi gas (conduction electrons)  and 
the Bose gas (magnons) coexist to interact.
The dynamics is described by 
the Hamiltonian  ${\cal{H}}_{\rm{ex}} ^{\prime}    $ and 
the width of the interface is  supposed to be
of the order of the lattice constant.\cite{interface}
Eq. (\ref{eqn:e3})  shows that localized spins at the interface lose  spin angular momentum by emitting  magnons
and 
conduction electrons flip from down to up by absorbing the spin angular momentum 
(see Fig. \ref{fig:pumping}), and vice versa.
This Hamiltonian ${\cal{H}}_{\rm{ex}} ^{\prime}    $, which describes the interchange of spin angular momentum
between localized spins and conduction electrons,     
is essential  to spin pumping mediated by magnons.

Therefore we clarify the contribution of magnons accompanying this exchange interaction to spin pumping,
under a time-dependent transverse magnetic field.
This is the main purpose of this paper. 
We also discuss how to enhance the spin pumping effect.

This paper is structured as follows.
First, through the  Heisenberg equation of motion,  the spin  relaxation  torque,
which breaks the spin conservation law for conduction electrons, 
is defined in \S  \ref{sec:source}. 
Second, we  evaluate it through the Schwinger-Keldysh formalism  in  \S  \ref{sec:kel}.
Third,  
we discuss  how to enhance the spin pumping effect mediated by magnons
through quantum fluctuations in \S  \ref{sec:pumping}.
Last, we  discuss the distinction between our quantum spin pumping theory and the one proposed by Tserkovnyak et al.\cite{mod2}
in \S \ref{sec:Tserkovnyak}. 

Let us remark that
in this paper,
we use the term, ${\textit{quantum fluctuations}}$,
to indicate applying time-dependent transverse magnetic fields.

\section{Spin Relaxation Torque}
\label{sec:source}

\subsection{Theoretical model}
\label{subsec:model}

We apply a time-dependent transverse magnetic field, which acts as a  quantum fluctuation; 
\begin{equation}
 \Gamma (t) \equiv  \Gamma _0 {\rm{cos}}(\Omega t).
 \end{equation}
 Then the  total Hamiltonian of the system (interface), ${\cal{H}}$,  is given as 
\begin{eqnarray}
{\cal{H}}  &=&   { \cal{H}}_{\rm{mag}} +  { \cal{H}}_{\rm{ex}}^{\prime} +  { \cal{H}}_{\rm{el}} + V_{\rm{el}}^{\Gamma }+  V_{\rm{mag}}^{\Gamma },  \  \rm{where}   \\
{\cal{H}}_{\rm{el}} &= &\int_{{\mathbf{x}}\in \text{(interface)}} d   {\mathbf{x}}   {\ }  c^{\dagger } ({\mathbf{x}},t) 
                                       \Big[  -\frac{\nabla ^2}{2m_{\rm{el}}}  -(JS + \frac{B}{2} )\sigma^z   \Big] c({\mathbf{x}},t). 
\label{eqn:e52}                                                          \\
 V_{\rm{el}}^{\Gamma }&=& \frac{\Gamma (t)}{4} \int_{{\mathbf{x}}\in \text{(interface)}} d   {\mathbf{x}}   {\ } 
                                                 c^{\dagger } ({\mathbf{x}},t) ( \sigma^{+}  +  \sigma^{-}  ) c({\mathbf{x}},t).  
\label{eqn:e53}     \\
 V_{\rm{mag}}^{\Gamma }&= &\Gamma (t)\sqrt{\frac{\tilde S}{2}} \int_{{\mathbf{x}}\in \text{(interface)}}  d   {\mathbf{x}}    {\ } \Bigg\{ 
                                                        \Big[1-\frac{a^{\dagger }({\mathbf{x}},t) a({\mathbf{x}},t) }{4\tilde S} \Big] a({\mathbf{x}},t)    \nonumber      \\
                                                &+&a^{\dagger }({\mathbf{x}},t) \Big[1-\frac{a^{\dagger }({\mathbf{x}},t) a({\mathbf{x}},t) }{4\tilde S} \Big]    \Bigg\}.
\label{eqn:e4}
\end{eqnarray}
The variable $m_{\rm{el}}$ denotes the effective mass of a conduction electron.

Here let us mention that in the last section,
we have rewritten  localized spin degrees of freedom into magnon ones via the Holstein-Primakoff transformation.
There, of course,  off-diagonal terms  (magnon-magnon interactions) may emerge as well as eq. (\ref{eqn:e2}),
but they cannot satisfy the resonance condition
discussed at \S  \ref{subsec:pump} in detail.
Consequently,
their contribution  to spin pumping 
is extremely smaller than $V_{\rm{mag}}^{\Gamma }$ (eq. (\ref{eqn:e4})) and negligible.
Therefore we omit such interactions   and 
concentrate on the effect of  $V_{\rm{mag}}^{\Gamma }$, which  represents
the coupling with the time-dependent transverse magnetic field. 

\subsection{Definition}
\label{subsec:def}

The spin relaxation torque (SRT),\cite{tsutsui}  ${\mathcal{T}}_{\rm{s}}^z$, is defined as the term 
which  breaks the spin conservation law for conduction electrons;\cite{gene,def}
 \begin{equation}
\begin{split}
 \dot \rho_{\rm{s}}^z  +   \nabla \cdot {\mathbf{j}}_{\rm{s}}^{z} =  {\mathcal{T}}_{\rm{s}}^z ,
\label{eqn:continuity}   
\end{split}
 \end{equation}
 where the dot denotes the time derivative,  ${\mathbf{j}}_{\rm{s}}$  is the spin current density, and
$\rho_{\rm{s}}^z$ represents the z-component of the spin density.
  We here have defined  the spin density of the system   as the expectation value (estimated for the total Hamiltonian, {$\cal{H}$});
\begin{eqnarray}   
  \rho_{\rm{s}}^z \equiv  \langle  c^{\dagger }  \sigma ^z c/2  \rangle.
  \end{eqnarray}   
In this paper, we focus on the z-component of the SRT.

Through the Heisenberg equation of motion,
the  z-component of the SRT   is defined as 
\begin{eqnarray}
{\mathcal{T}}_{\rm{s}}^z  &=& \bigg{\langle}    iJa_0^3  \sqrt{\frac{\tilde S}{2}}  \Big{\{} a^{\dagger }({\mathbf{x}},t) 
                                                         \Big[1-\frac{ a^{\dagger }({\mathbf{x}},t) a({\mathbf{x}},t)}{4 \tilde S }\Big]      c^{\dagger }  ({\mathbf{x}},t) \sigma^{+}   c({\mathbf{x}},t ) \nonumber \\
                                             & - &  \Big[1-\frac{ a^{\dagger }({\mathbf{x}},t) a({\mathbf{x}},t)}{4 \tilde S }\Big]a({\mathbf{x}},t)  
                                                          c^{\dagger } ({\mathbf{x}},t)  \sigma^{-}   c ({\mathbf{x}},t)  \Big{\}}    \nonumber  \\
                                             &+ & \frac{\Gamma (t)}{4i}  
                                                       \Big[ c^{\dagger }   ({\mathbf{x}},t) \sigma^{+}   c({\mathbf{x}},t )    -    c^{\dagger } ({\mathbf{x}},t)  \sigma^{-}   c ({\mathbf{x}},t) \Big]  \bigg{\rangle}.
\label{eqn:e5}                       
\end{eqnarray}
This term arises from 
  $ { \cal{H}}_{\rm{ex}}^{\prime} $ and $ V_{\rm{el}}^{\Gamma }$,
which consist of electron spin-flip  operators;   
\begin{eqnarray}   
 {\mathcal{T}}_{\rm{s}}^z  =   [\rho_{\rm{s}}^z,  { \cal{H}}_{\rm{ex}}^{\prime}    +  V_{\rm{el}}^{\Gamma }  ]/i.
 \end{eqnarray}   
  Eq. (\ref{eqn:e5}) shows that the SRT operates the coherent magnon state.\cite{mista}

 According to Ralph et al., \cite{torqueJMM}
 the net flux of nonequilibrium  spin current (i.e. the net spin current) pumped through the surface of the interface,
 $\int      \    {\mathbf{j}}_{\rm{s}}^{z}     \cdot        d {\mathbf{S}}_{\rm{interface}}         $,
can be computed simply by integrating the SRT over the  volume of the interface;
\begin{eqnarray}   
 \int_{{\mathbf{x}}\in \rm{(interface)}}    d{\mathbf{x}}    \   {\mathcal{T}}_{\rm{s}}^z  
= \int     \    {\mathbf{j}}_{\rm{s}}^{z}         \cdot        d {\mathbf{S}}_{\rm{interface}}.
\end{eqnarray}   
 In addition in our case, conduction electrons  cannot  enter the ferromagnet, which is an insulator.\cite{spinwave} 
Therefore  as shown in  Fig. \ref{fig:pumping} (b), 
the z-component of the net spin current pumped into non-magnetic metal  can be expressed
as 
\begin{eqnarray}   
 \int_{{\mathbf{x}}\in \rm{(interface)}}    d{\mathbf{x}}    \   {\mathcal{T}}_{\rm{s}}^z.
 \end{eqnarray}   

Thus from now on, we focus on  $ {\mathcal{T}}_{\rm{s}}^z$ 
and 
qualitatively  clarify the behavior of the spin pumping effect mediated by magnons 
at  room temperature.
Let us mention that 
the above relation between the SRT and the pumped net spin current can be understood 
via the spin continuity equation, eq. (\ref{eqn:continuity}),
 and
we discuss  at \S  \ref{subsec:conservation}.

\subsection{Magnon continuity equation}
\label{subsec:mag}

Here let us emphasize that 
the spin conservation law for localized spins (i.e. magnons) is also broken.
The magnon continuity equation for localized spins,  \cite{nakatatatara} 
which corresponds to the equation of motion for localized spins\cite{tataraprivate}
and describes the dynamics,
reads 
\begin{eqnarray}
\dot \rho_{\rm{m}}^z  +   \nabla \cdot {\mathbf{j}}_{\rm{m}}^{z} =  {\mathcal{T}}_{\rm{m}}^z ,
\label{eqn:e554}  
\end{eqnarray}
 where ${\mathbf{j}}_{\rm{m}}$  is the magnon current density, and
$\rho_{\rm{m}}^z$ represents the z-component of the magnon density.
 We  have  defined  the magnon density of the system also  as the expectation value (estimated for the total Hamiltonian, {$\cal{H}$});
 \begin{eqnarray}
 \rho_{\rm{m}}^z \equiv  \langle  a^{\dagger } a  \rangle.
 \end{eqnarray}
 In addition, we call ${\mathcal{T}}_{\rm{m}}^z$ the magnon source term  \cite{nakatatatara},
which breaks the magnon conservation law.
This term arises  also from ${\cal{H}}_{\rm{ex}}^{\prime}$ and $V_{\rm{mag}}^{\Gamma }$;
 \begin{eqnarray}
 {\mathcal{T}}_{\rm{m}}^z &=& [\rho_{\rm{m}}^z, {\cal{H}}_{\rm{ex}}^{\prime} + V_{\rm{mag}}^{\Gamma }]/i           \\   
 &=& \Bigg{\langle}    iJa_0^3  \sqrt{\frac{\tilde S}{2}}  \Bigg{\{} a^{\dagger }({\mathbf{x}},t) 
                                                         \Big[1-\frac{ a^{\dagger }({\mathbf{x}},t) a({\mathbf{x}},t)}{4 \tilde S }\Big]      c^{\dagger }  ({\mathbf{x}},t) \sigma^{+}   c({\mathbf{x}},t ) \nonumber \\
                                             & - &  \Big[1-\frac{ a^{\dagger }({\mathbf{x}},t) a({\mathbf{x}},t)}{4 \tilde S }\Big]a({\mathbf{x}},t)  
                                                          c^{\dagger } ({\mathbf{x}},t)  \sigma^{-}   c ({\mathbf{x}},t)  \Bigg{\}}    \nonumber  \\
                                             &+ &  i\Gamma (t)  \sqrt{\frac{\tilde S}{2}} 
                                                       \Bigg\{ \Big[1-\frac{ a^{\dagger }({\mathbf{x}},t) a({\mathbf{x}},t)}{4 \tilde S }\Big]a({\mathbf{x}},t)  
                                                                        -  a^{\dagger }({\mathbf{x}},t) 
                                                         \Big[1-\frac{ a^{\dagger }({\mathbf{x}},t) a({\mathbf{x}},t)}{4 \tilde S }\Big]  
                                                              \Bigg\}  \Bigg{\rangle}.
 \end{eqnarray}

Within the same approximation with the SRT (, which is discussed at the next section in detail), 
this magnon source term, in fact, satisfies the relation;
\begin{eqnarray}
 {\mathcal{T}}_{\rm{m}}^z = {\mathcal{T}}_{\rm{s}}^z.
 \label{eqn:e61}  
 \end{eqnarray}
Then, the z-component of the spin continuity equation for the total system (i.e. conduction electrons and magnons) becomes
\begin{eqnarray}
\dot {\rho} _{\rm{total}}^z + \nabla \cdot  {\mathbf{j}}_{\rm{total}}^z =0,
\label{eqn:e555}  
\end{eqnarray}
where the density of the total spin angular momentum, $ {\rho} _{\rm{total}}^z$, is defined as
\begin{eqnarray}
{\rho} _{\rm{total}}^z \equiv  {\rho} _{\rm{s}}^z -  {\rho} _{\rm{m}}^z,
\end{eqnarray}
and consequently
the z-component of the total spin current density,  ${\mathbf{j}}_{\rm{total}}^z$, becomes
\begin{eqnarray}
 {\mathbf{j}}_{\rm{total}}^z  =    {\mathbf{j}}_{\rm{s}}^z -  {\mathbf{j}}_{\rm{m}}^z
 \end{eqnarray}
 (note that, $S^z = \tilde S - a^{\dagger } a$,  via the Holstein-Primakoff transformation).
The spin continuity equation for the whole system, eq. (\ref{eqn:e555}),
 means that 
though each spin conservation law for electrons and magnons is broken (see eqs. (\ref{eqn:continuity}) and (\ref{eqn:e554})),
the total spin angular momentum  is, of course, conserved.

\section{Schwinger-Keldysh Formalism}
\label{sec:kel}

The interface  is,  in general,   a weak coupling regime;\cite{AndoPumping}
the exchange interaction, $J$, is supposed to be
 smaller than the Fermi energy and
the exchange interaction among ferromagnets.
Thus  $ { \cal{H}}_{\rm{ex}}^{\prime} $ can be treated as a perturbative term.
In addition,
we apply weak transverse magnetic fields.
Then 
we can treat $ { \cal{H}}_{\rm{ex}}^{\prime} $,   $V_{\rm{el}}^{\Gamma } $, and   $V_{\rm{mag}}^{\Gamma } $
as perturbative terms
to evaluate the SRT, eq. (\ref{eqn:e5}).

Through the standard procedure of the Schwinger-Keldysh (or contour-ordered) Green's function (see also Appendix \ref{subsec:boson}),\cite{ramer,kamenev,kita}
the Langreth method (see also Appendix \ref{subsec:concrete}),\cite{haug,tatara,new,rammer}
each term of  the SRT  can be  evaluated as follows;
the first term of eq. (\ref{eqn:e5}) reads (see also Appendix \ref{subsec:ours})
\begin{equation}
\begin{split}
\bigg{\langle}    iJa_0^3  \sqrt{\frac{\tilde S}{2}}   a^{\dagger }({\mathbf{x}},t) 
                                                    \Big[1-\frac{ a^{\dagger }({\mathbf{x}},t) a({\mathbf{x}},t)}{4 \tilde S }\Big]
                                                      c^{\dagger }  ({\mathbf{x}},t) \sigma^{+}   c({\mathbf{x}},t )  \bigg{\rangle}      {\            \            \          \                    }    \\
             =    \frac{J S}{2}         (\frac{\Gamma _0}{2})^2       \int \frac{ d   {\mathbf{k}} }{ (2\pi)^3} \int \frac{d\omega}{2\pi}   
                      \Big[1-\frac{i}{\tilde S }    \int \frac{ d{    {\mathbf{k}}^{\prime}}}{ (2\pi)^3}   \int \frac{d{\omega}^{\prime} }{2\pi} 
                                 {\rm{G}}^{\rm{<}}_{{\mathbf{k^{\prime}}}, \omega^{\prime} } \Big]    \\   
     \times  \Bigg\{  \Big[  {\rm{e}}^{2i\Omega t} {\rm{G}}^{\rm{a}}_{0, \Omega } + {\rm{G}}^{\rm{a}}_{0, -\Omega }   \Big]
                                    \Big[   {\mathcal{G}}^{\rm{t}}_{\downarrow , {\mathbf{k}}, \omega -\Omega } 
                                       {\mathcal{G}}^{\rm{t}}_{\uparrow  , {\mathbf{k}}, \omega  }  
                                    -   {\mathcal{G}}^{\rm{<}}_{\downarrow , {\mathbf{k}}, \omega -\Omega } 
                                       {\mathcal{G}}^{\rm{>}}_{\uparrow  , {\mathbf{k}}, \omega  } \Big]        {\             \             }      \\
       +     \Big[  {\rm{e}}^{-2i\Omega t} {\rm{G}}^{\rm{a}}_{0, -\Omega } + {\rm{G}}^{\rm{a}}_{0, \Omega }   \Big]
                                   \Big[    {\mathcal{G}}^{\rm{t}}_{\downarrow , {\mathbf{k}}, \omega +\Omega } 
                                       {\mathcal{G}}^{\rm{t}}_{\uparrow  , {\mathbf{k}}, \omega  }  
                                   -    {\mathcal{G}}^{\rm{<}}_{\downarrow , {\mathbf{k}}, \omega +\Omega } 
                                       {\mathcal{G}}^{\rm{>}}_{\uparrow  , {\mathbf{k}}, \omega  }   \Big]                         
                     \Bigg\}  \\
        +   {\cal{O}}(J^2)   + {\cal{O}}(\Gamma ^4) +   {\cal{O}}(JS^{-1}).      {\               \                     \          \        \              }    
\label{eqn:e6}
\end{split}
\end{equation}
The variable   $  {\mathcal{G}}^{{\rm{t}} (<, >)} $  
is the fermionic time-ordered (lesser, greater)  Green's function,
and $  {\rm{G}}^{\rm{<} (\rm{a})}$ is the bosonic lesser (advanced) one.

We here have  taken the extended time (i.e. the contour variable) defined on the  Schwinger-Keldysh closed time path,\cite{rammer,kita,tatara,haug,new} c, 
on the forward path ${\text{c}}_{\rightarrow }$ (see also Fig. A$\cdot $1 in our manuscript\cite{nakatatatara}); 
$ c =  {\text{c}}_{\rightarrow } + {\text{c}}_{\leftarrow } $.
Even when the time is located  on the backward path ${\text{c}}_{\leftarrow }$, 
the result of the calculation does not change
because each Green's function is not independent;
$ {\mathcal{G}}^{\text{r}}  - {\mathcal{G}}^{\text{a}}  = {\mathcal{G}}^{\text{>}}  - {\mathcal{G}}^{\text{<}} $,
where $ {\mathcal{G}}^{\text{r}(\text{a})}$ represents the  retarded (advanced) Green's function.
This relation comes into effect also for the bosonic case (see eq. (\ref{eqn:appe10})).  

Here it would be useful to mention that
under the thermal equilibrium condition  where  temperature difference  does not exist  between ferromagnet and non-magnetic metal,
the  $ {\cal{O}}(J^2 \Gamma ^0)$ term including no quantum fluctuations  
cannot contribute to spin pumping
because of the balance between thermal fluctuations in ferromagnet and those in non-magnetic metal.\cite{adachiseebeck}

The second term of eq. (\ref{eqn:e5}) reads (see also Appendix \ref{subsec:ours})
\begin{equation}
\begin{split}
\bigg{\langle}    -iJa_0^3  \sqrt{\frac{\tilde S}{2}}    \Big[1-\frac{ a^{\dagger }({\mathbf{x}},t) a({\mathbf{x}},t)}{4 \tilde S }\Big] a({\mathbf{x}},t) 
                                                      c^{\dagger }  ({\mathbf{x}},t) \sigma^{-}   c({\mathbf{x}},t )  \bigg{\rangle}   { \    \          \        }   \\
             =    - \frac{J S}{2}         (\frac{\Gamma _0}{2})^2      \int \frac{ d    {\mathbf{k}} }{ (2\pi)^3}   \int \frac{d\omega}{2\pi}   
                       \Big[1-\frac{i}{\tilde S }    \int \frac{ d{  {\mathbf{k}}^{\prime}}}{ (2\pi)^3}   \int \frac{d{\omega}^{\prime} }{2\pi} 
                                 {\rm{G}}^{\rm{<}}_{{\mathbf{k^{\prime}}}, \omega^{\prime} } \Big]      \\   
     \times  \Bigg\{  \Big[  {\rm{e}}^{2i\Omega t} {\rm{G}}^{\rm{r}}_{0, \Omega } + {\rm{G}}^{\rm{r}}_{0, -\Omega }   \Big]
                                    \Big[   {\mathcal{G}}^{\rm{t}}_{\uparrow , {\mathbf{k}}, \omega -\Omega } 
                                       {\mathcal{G}}^{\rm{t}}_{\downarrow  , {\mathbf{k}}, \omega  }  
                                    -   {\mathcal{G}}^{\rm{<}}_{\uparrow , {\mathbf{k}}, \omega -\Omega } 
                                       {\mathcal{G}}^{\rm{>}}_{\downarrow  , {\mathbf{k}}, \omega  }  \Big]         {\            \               \                       }   \\
       +     \Big[  {\rm{e}}^{-2i\Omega t} {\rm{G}}^{\rm{r}}_{0, -\Omega } + {\rm{G}}^{\rm{r}}_{0, \Omega }   \Big]
                                 \Big[      {\mathcal{G}}^{\rm{t}}_{\uparrow , {\mathbf{k}}, \omega +\Omega } 
                                       {\mathcal{G}}^{\rm{t}}_{\downarrow  , {\mathbf{k}}, \omega  }  
                                   -    {\mathcal{G}}^{\rm{<}}_{\uparrow , {\mathbf{k}}, \omega +\Omega } 
                                       {\mathcal{G}}^{\rm{>}}_{\downarrow  , {\mathbf{k}}, \omega  }     \Big]                       
                     \Bigg\}        {\                   }         \\
           +     {\cal{O}}(J^2)   + {\cal{O}}(\Gamma ^4)+   {\cal{O}}(JS^{-1}),     {\               \                     \                 } 
\label{eqn:e7}
\end{split}
\end{equation}
where the variable  $  {\rm{G}}^{\rm{r} }$ is the bosonic retarded  Green's function.
The third term of eq. (\ref{eqn:e5}) reads (see also Appendix \ref{subsec:ours})
\begin{equation}
\begin{split}
\bigg{\langle}   \frac{\Gamma (t)  c^{\dagger }  ({\mathbf{x}},t) \sigma^{+}   c({\mathbf{x}},t )  }{4i}  \bigg{\rangle}  
                {\         \               \              \                \              \                   \                  \                      \                            \         \               \              \                \              \              
             \                  \                      \         \         \               \              \            \                    \                       \                     \              \                  \                      \     \  
                    \                    \                       \                     \              \                  \                      \     \       \                  \                 \                \                 \                    \         \   }            \\
             =     \frac{J S}{2}         (\frac{\Gamma _0}{2})^2       \int \frac{ d  {\mathbf{k}} }{ (2\pi)^3}     \int \frac{d\omega}{2\pi}      
       \Big[1-\frac{i}{\tilde S }     \int \frac{ d{ {\mathbf{k}}^{\prime}}}{ (2\pi)^3}   \int \frac{d{\omega}^{\prime} }{2\pi} 
                                 {\rm{G}}^{\rm{<}}_{{\mathbf{k^{\prime}}}, \omega^{\prime} } \Big]       \\  
   \times   \Bigg\{     ({\rm{e}}^{2i\Omega t}+1) {\rm{G}}^{\rm{r}}_{0, -\Omega } 
                                     \Big[    {\mathcal{G}}^{\rm{t}}_{\downarrow , {\mathbf{k}}, \omega -\Omega } 
                                       {\mathcal{G}}^{\rm{t}}_{\uparrow  , {\mathbf{k}}, \omega  }  
                                    -   {\mathcal{G}}^{\rm{<}}_{\downarrow , {\mathbf{k}}, \omega -\Omega } 
                                       {\mathcal{G}}^{\rm{>}}_{\uparrow  , {\mathbf{k}}, \omega  }   \Big]     {\       \         \         \      \ }    \\
       +      ( {\rm{e}}^{-2i\Omega t}+1) {\rm{G}}^{\rm{r}}_{0, \Omega } 
                                    \Big[  {\mathcal{G}}^{\rm{t}}_{\downarrow , {\mathbf{k}}, \omega +\Omega } 
                                       {\mathcal{G}}^{\rm{t}}_{\uparrow  , {\mathbf{k}}, \omega  }  
                                   -    {\mathcal{G}}^{\rm{<}}_{\downarrow , {\mathbf{k}}, \omega +\Omega } 
                                       {\mathcal{G}}^{\rm{>}}_{\uparrow  , {\mathbf{k}}, \omega  }     \Big]                       
                     \Bigg\}       {\       \         \         \      \ }            \\
            +  {\cal{O}}(J^0)  + {\cal{O}}(J^2)   + {\cal{O}}(\Gamma ^4)+   {\cal{O}}(JS^{-1}).     {\               \                     \          \        \              } 
\label{eqn:e8}
\end{split}
\end{equation}
We omit the $   {\cal{O}}(J^0) $ terms
because they contain no contributions of magnons via the exchange interaction
and  are not relevant to the spin pumping effect;
the terms are out of the purpose of  the present study.

The last term of eq. (\ref{eqn:e5}) reads (see also Appendix \ref{subsec:ours})
\begin{equation}
\begin{split}
\bigg{\langle}  - \frac{\Gamma (t)  c^{\dagger }  ({\mathbf{x}},t) \sigma^{-}   c({\mathbf{x}},t )  }{4i}  \bigg{\rangle}   
 {\             \           \           \          \             \               \                \                \                  \          \             \           \           \          \             \               \                \                \                  \ 
 \             \           \           \          \             \               \                \                \                  \          \             \           \           \               \              \                     \
          \                        \               \                     \                     \                   \                        \ }      \\
             =   -  \frac{J S}{2}         (\frac{\Gamma _0}{2})^2       \int \frac{ d    {\mathbf{k}}}{ (2\pi)^3}   \int \frac{d\omega}{2\pi}    
                    \Big[1-\frac{i}{\tilde S }   \int \frac{ d{ {\mathbf{k}}^{\prime}}}{ (2\pi)^3}    \int \frac{d{\omega}^{\prime} }{2\pi} 
                                 {\rm{G}}^{\rm{<}}_{{\mathbf{k^{\prime}}}, \omega^{\prime} } \Big]           \\     
   \times    \Bigg\{     ({\rm{e}}^{2i\Omega t}+1) {\rm{G}}^{\rm{a}}_{0, -\Omega } 
                                     \Big[    {\mathcal{G}}^{\rm{t}}_{\uparrow , {\mathbf{k}}, \omega -\Omega } 
                                       {\mathcal{G}}^{\rm{t}}_{\downarrow  , {\mathbf{k}}, \omega  }  
                                    -   {\mathcal{G}}^{\rm{<}}_{\uparrow , {\mathbf{k}}, \omega -\Omega } 
                                       {\mathcal{G}}^{\rm{>}}_{\downarrow  , {\mathbf{k}}, \omega  }   \Big]    {\       \         \         \      \ }    \\
       +      ( {\rm{e}}^{-2i\Omega t}+1) {\rm{G}}^{\rm{a}}_{0, \Omega } 
                                    \Big[  {\mathcal{G}}^{\rm{t}}_{\uparrow , {\mathbf{k}}, \omega +\Omega } 
                                       {\mathcal{G}}^{\rm{t}}_{\downarrow  , {\mathbf{k}}, \omega  }  
                                   -    {\mathcal{G}}^{\rm{<}}_{\uparrow , {\mathbf{k}}, \omega +\Omega } 
                                       {\mathcal{G}}^{\rm{>}}_{\downarrow  , {\mathbf{k}}, \omega  }     \Big]                       
                     \Bigg\}     {\       \         \         \      \ }       \\
             +  {\cal{O}}(J^0) + {\cal{O}}(J^2)   + {\cal{O}}(\Gamma ^4)+   {\cal{O}}(JS^{-1}).   {\               \                     \          \        \              }        
\label{eqn:e9}
\end{split}
\end{equation}

Finally, 
the SRT can be rearranged as 
\begin{equation}
\begin{split}
{\mathcal{T}}_{\rm{s}}^z  =  \sum_{n=0, \pm 1}  {\mathcal{T}}_{\rm{s}}^{z} (n)   \   {\rm{e}}^{2in\Omega t},   
\label{eqn:e20}   
\end{split}                          
\end{equation}
where
\begin{eqnarray}
{\mathcal{T}}_{\rm{s}}^z (1) &=&                  \frac{J S}{2}         (\frac{\Gamma _0}{2})^2       \int \frac{ d  {\mathbf{k}} }{ (2\pi)^3}     \int \frac{d\omega}{2\pi}    
                                                         \Big[1-\frac{i}{\tilde S }     \int \frac{ d{ {\mathbf{k}}^{\prime}}}{ (2\pi)^3}   \int \frac{d{\omega}^{\prime} }{2\pi} 
                                                                                                                                         {\rm{G}}^{\rm{<}}_{{\mathbf{k^{\prime}}}, \omega^{\prime} } \Big]     \nonumber  \\
                                                  &\times  &     \Big[({\rm{G}}^{\rm{a}}_{0, \Omega }  +   {\rm{G}}^{\rm{r}}_{0, -\Omega })    
                                                           ({\mathcal{G}}^{\rm{t}}_{\downarrow , {\mathbf{k}}, \omega -\Omega } 
                                                                                {\mathcal{G}}^{\rm{t}}_{\uparrow  , {\mathbf{k}}, \omega  }  
                                                                              -   {\mathcal{G}}^{\rm{<}}_{\downarrow , {\mathbf{k}}, \omega -\Omega } 
                                                                                {\mathcal{G}}^{\rm{>}}_{\uparrow  , {\mathbf{k}}, \omega  })                    \nonumber         \\
                                                 &- &                       ({\rm{G}}^{\rm{r}}_{0, \Omega }  +  {\rm{G}}^{\rm{a}}_{0, -\Omega } )  
                                                             (   {\mathcal{G}}^{\rm{t}}_{\uparrow , {\mathbf{k}}, \omega -\Omega } 
                                                                                {\mathcal{G}}^{\rm{t}}_{\downarrow  , {\mathbf{k}}, \omega  }  
                                                                                   -   {\mathcal{G}}^{\rm{<}}_{\uparrow , {\mathbf{k}}, \omega -\Omega } 
                                                                                  {\mathcal{G}}^{\rm{>}}_{\downarrow  , {\mathbf{k}}, \omega  }  )\Big],  \label{eqn:e71}                                                                               
 \end{eqnarray}
\begin{eqnarray}
{\mathcal{T}}_{\rm{s}}^z (-1)     &=&                     \frac{J S}{2}         (\frac{\Gamma _0}{2})^2       \int \frac{ d  {\mathbf{k}} }{ (2\pi)^3}     \int \frac{d\omega}{2\pi}    
                                                              \Big[1-\frac{i}{\tilde S }     \int \frac{ d{ {\mathbf{k}}^{\prime}}}{ (2\pi)^3}   \int \frac{d{\omega}^{\prime} }{2\pi} 
                                                                                                                                         {\rm{G}}^{\rm{<}}_{{\mathbf{k^{\prime}}}, \omega^{\prime} } \Big]     \nonumber  \\
                                                       &\times &          \Big[( {\rm{G}}^{\rm{a}}_{0, -\Omega }  +{\rm{G}}^{\rm{r}}_{0, \Omega } )  
                                                                  (    {\mathcal{G}}^{\rm{t}}_{\downarrow , {\mathbf{k}}, \omega +\Omega } 
                                                                                         {\mathcal{G}}^{\rm{t}}_{\uparrow  , {\mathbf{k}}, \omega  }  
                                                                                           - {\mathcal{G}}^{\rm{<}}_{\downarrow , {\mathbf{k}}, \omega +\Omega } 
                                                                                            {\mathcal{G}}^{\rm{>}}_{\uparrow  , {\mathbf{k}}, \omega  }   )     \nonumber  \\
                                                      &-  &                      ( {\rm{G}}^{\rm{r}}_{0, -\Omega } +    {\rm{G}}^{\rm{a}}_{0, \Omega }  )  
                                                                  (     {\mathcal{G}}^{\rm{t}}_{\uparrow , {\mathbf{k}}, \omega +\Omega } 
                                                                                            {\mathcal{G}}^{\rm{t}}_{\downarrow  , {\mathbf{k}}, \omega  }  
                                                                                            -    {\mathcal{G}}^{\rm{<}}_{\uparrow , {\mathbf{k}}, \omega +\Omega } 
                                                                                         {\mathcal{G}}^{\rm{>}}_{\downarrow  , {\mathbf{k}}, \omega  }     )\Big],    
\label{eqn:e72}
   \end{eqnarray}
\begin{eqnarray}
  {\mathcal{T}}_{\rm{s}}^z (0)   &=  &                     \frac{J S}{2}         (\frac{\Gamma _0}{2})^2       \int \frac{ d  {\mathbf{k}} }{ (2\pi)^3}     \int \frac{d\omega}{2\pi}    
                                                           \Big[1-\frac{i}{\tilde S }     \int \frac{ d{ {\mathbf{k}}^{\prime}}}{ (2\pi)^3}   \int \frac{d{\omega}^{\prime} }{2\pi} 
                                                                                                                                         {\rm{G}}^{\rm{<}}_{{\mathbf{k^{\prime}}}, \omega^{\prime} } \Big]     \nonumber  \\
                                                    &\times  &            \Big[ ({\rm{G}}^{\rm{a}}_{0, -\Omega }   +  {\rm{G}}^{\rm{r}}_{0, -\Omega }  )   
                                                              (   {\mathcal{G}}^{\rm{t}}_{\downarrow , {\mathbf{k}}, \omega -\Omega } 
                                                                                   {\mathcal{G}}^{\rm{t}}_{\uparrow  , {\mathbf{k}}, \omega  }  
                                                                                      -   {\mathcal{G}}^{\rm{<}}_{\downarrow , {\mathbf{k}}, \omega -\Omega } 
                                                                                     {\mathcal{G}}^{\rm{>}}_{\uparrow  , {\mathbf{k}}, \omega  })       \nonumber    \\
                                                  &+   &                      ( {\rm{G}}^{\rm{a}}_{0, \Omega }  +  {\rm{G}}^{\rm{r}}_{0, \Omega }  )  
                                                                 (  {\mathcal{G}}^{\rm{t}}_{\downarrow , {\mathbf{k}}, \omega +\Omega } 
                                                                                        {\mathcal{G}}^{\rm{t}}_{\uparrow  , {\mathbf{k}}, \omega  }  
                                                                                         -    {\mathcal{G}}^{\rm{<}}_{\downarrow , {\mathbf{k}}, \omega +\Omega } 
                                                                                           {\mathcal{G}}^{\rm{>}}_{\uparrow  , {\mathbf{k}}, \omega  }  )  \nonumber \\
                                                  &-  &                        ({\rm{G}}^{\rm{r}}_{0, -\Omega }  +{\rm{G}}^{\rm{a}}_{0, -\Omega } )   
                                                                       (  {\mathcal{G}}^{\rm{t}}_{\uparrow , {\mathbf{k}}, \omega -\Omega } 
                                                                                           {\mathcal{G}}^{\rm{t}}_{\downarrow  , {\mathbf{k}}, \omega  }  
                                                                                           -   {\mathcal{G}}^{\rm{<}}_{\uparrow , {\mathbf{k}}, \omega -\Omega } 
                                                                                          {\mathcal{G}}^{\rm{>}}_{\downarrow  , {\mathbf{k}}, \omega  } )   \nonumber  \\
                                                 &- &                           ( {\rm{G}}^{\rm{r}}_{0, \Omega }   +{\rm{G}}^{\rm{a}}_{0, \Omega } )  
                                                                  (      {\mathcal{G}}^{\rm{t}}_{\uparrow , {\mathbf{k}}, \omega +\Omega } 
                                                                                            {\mathcal{G}}^{\rm{t}}_{\downarrow  , {\mathbf{k}}, \omega  }  
                                                                                          -    {\mathcal{G}}^{\rm{<}}_{\uparrow , {\mathbf{k}}, \omega +\Omega } 
                                                                                           {\mathcal{G}}^{\rm{>}}_{\downarrow  , {\mathbf{k}}, \omega  } )\Big].
\label{eqn:e21}                          
\end{eqnarray}

Here let us mention that,
in real materials, there does exist  impurity scattering.
We assume that this is the main cause for  the finite lifetime  of magnons and conduction electrons. 
Moreover
the rate of impurities such as lattice defects and nonmagnetic impurities is, in general, far larger than that of magnetic impurities. 
Therefore we  phenomenologically introduce the lifetime  and regard it as  a constant parameter.
Then we adopt Green's functions including the the effects of impurities as the lifetime,
and calculate the SRT by using them; eqs. (\ref{eqn:G1})-(\ref{eqn:G8}).

Though, as the result, 
it might be better to execute the accompanying vertex corrections from viewpoints of theoretical aspects,
we have not done for the aim now explained;
to put it briefly,  
in order to clarify that pumped spin currents are generated purely by quantum fluctuations,
we have not executed vertex corrections.

It should be noted that, before our present study,
Takeuchi et al.\cite{takeuchi} have already studied spin pumping, on the basis of Schwinger-Keldysh formalism,
under the same condition with ours except two points; 
(i) they have treated localized spins as not  magnons but classical variables, and 
(ii) they have not applied any transverse magnetic fields.
On their condition,
they have clarified that, under the uniform magnetization,  
spin currents cannot be generated without vertex corrections (i.e. multiple scatterings of impurities).
In other words, they have already  revealed that 
spin currents can be generated by the effects of multiple scatterings of impurities, i.e. vertex corrections.   

Thus, the main purpose of the present study is 
to propose an alternative way for the generation of spin currents 
without using  vertex corrections, i.e. multiple scatterings of impurities;
we propose a method for the generation of spin currents
by using time-dependent transverse magnetic fields,
which are under our control and act as quantum fluctuations.
Therefore we call this method quantum spin pumping.  
In order to clarify that  pumped spin currents are induced purely by quantum fluctuations,
we have not included the effects of  multiple scatterings of impurities
(i.e. we have not executed vertex corrections).

Each Green's function including the the effects of impurities as the lifetime
reads as follows (see also Appendix \ref{subsec:boson});\cite{kamenev}
\begin{eqnarray}
 {\rm{G}}^{\rm{a}}_{   {\mathbf{k^{\prime}}}, \omega^{\prime} } &=& [\omega^{\prime} -  \omega_{\mathbf{k^{\prime}}} - i/(2\tau_{\rm{m}}) ]^{-1}    \label{eqn:G1}   \\ 
&=&  ({\rm{G}}^{\rm{r}}_{   {\mathbf{k^{\prime}}}, \omega^{\prime} })^{\ast },       \label{eqn:G2}     \\
 {\mathcal{G}}_{\sigma, {\mathbf{k}}, \omega}^{\rm{a}} &=& [\omega -  \omega_{\sigma , \mathbf{k}} - i/(2\tau) ]^{-1}   \label{eqn:G3}    \\  
                                &=&   ({\mathcal{G}}_{\sigma, {\mathbf{k}}, \omega}^{\rm{r}})^{\ast },     \label{eqn:G4}        \\
{\rm{G}}^{\rm{<}}_{   {\mathbf{k^{\prime}}}, \omega^{\prime} }  &=& -f_{\rm{B}}(\omega^{\prime}) 
                                   ( {\rm{G}}^{\rm{a}}_{   {\mathbf{k^{\prime}}}, \omega^{\prime} }-  {\rm{G}}^{\rm{r}}_{   {\mathbf{k^{\prime}}}, \omega^{\prime} }), \label{eqn:G5}    \\
{\mathcal{G}}_{\sigma, {\mathbf{k}}, \omega}^{\rm{<}}  &=& f_{\rm{F}} (\omega )
                                                                  ( {\mathcal{G}}_{\sigma, {\mathbf{k}}, \omega}^{\rm{a}}    -    {\mathcal{G}}_{\sigma, {\mathbf{k}}, \omega}^{\rm{r}}), \label{eqn:G6}   \\
{\mathcal{G}}_{\sigma, {\mathbf{k}}, \omega}^{\rm{t}}  &=&
                    {\mathcal{G}}_{\sigma, {\mathbf{k}}, \omega}^{\rm{r}}  +  {\mathcal{G}}_{\sigma, {\mathbf{k}}, \omega}^{\rm{<}}     \label{eqn:G7}      \\
                                                 &=&{\mathcal{G}}_{\sigma, {\mathbf{k}}, \omega}^{\rm{>}}  +    {\mathcal{G}}_{\sigma, {\mathbf{k}}, \omega}^{\rm{a}}, \label{eqn:G8}  
\end{eqnarray}                                              
where the variables
$\tau_{\rm{m}}$, $\tau$, $f_{\rm_{B}} (\omega^{\prime} )$, and  $f_{\rm_{F}} (\omega )$  are 
the lifetime of magnons, that of conduction electrons, the Bose distribution function, and the Fermi one, respectively.  
The energy dispersion relation reads
 $  \omega _{\mathbf{k^{\prime}}} \equiv  D {k^{\prime}}^2+B $ and 
$  \omega _{\sigma, \mathbf{k}} \equiv  F k^2- ( JS + B/2 )\sigma - \mu $,
where  $ D   \equiv  1/(2m)    $, $  F\equiv    1/(2m_{\rm{el}})  $,  $ \sigma = +1, -1 (= \uparrow, \downarrow )$, 
and $\mu$  denotes  the chemical potential.

We consider a weak magnetic field regime and omit the  $ {\cal{O}}([(JS+B)/\epsilon _{\rm{F}}]^2) $ terms,  
where $ \epsilon _{\rm{F}}$ represents the Fermi energy.
Through the Sommerfeld expansion, the chemical potential is determined as,
$ \mu (T) =  \epsilon _{\rm{F}}  - {(\pi k_{\rm{B} }T)^2}/({12\epsilon _{\rm{F}}}) + {\cal{O}}(T^4)$.  

\section{Spin Pumping}
\label{sec:pumping}

\subsection{Breaking of  spin conservation law }
\label{subsec:conservation}

The relation  between the SRT and the pumped net spin current in \S  \ref{subsec:def},
$ \int_{{\mathbf{x}}\in \rm{(interface)}}    d{\mathbf{x}}    \   {\mathcal{T}}_{\rm{s}}^z  
= \int     \  {\mathbf{j}}_{\rm{s}}^{z} \cdot        d {\mathbf{S}}_{\rm{interface}}       $,
can be understood  via the spin continuity equation; 
$  \dot \rho_{\rm{s}}^z  +   \nabla \cdot {\mathbf{j}}_{\rm{s}}^{z} =  {\mathcal{T}}_{\rm{s}}^z  $.

Through the same procedure and the same approximation with the last section,  
the time derivative of the spin density is estimated   
(on the condition mentioned at  \S \ref{subsec:pump}) as,
$ {\dot \rho_{\rm{s}}^z} /   {{\mathcal{T}}_{\rm{s}}^z} \sim     10^{-3}   $.
Thus  $ \dot \rho_{\rm{s}}^z$  in the spin continuity equation
 is negligible\cite{adachiseebeck} also in our case.    

Consequently,
 the spin continuity equation becomes
$ \nabla \cdot {\mathbf{j}}_{\rm{s}}^{z} =  {\mathcal{T}}_{\rm{s}}^z$.
Therefore by integrating  over the volume of the interface,
the relation between the SRT and the pumped net spin current mediated by magnons is clarified;
 \begin{eqnarray}
                   \int_{{\mathbf{x}}\in \rm{(interface)}}    d{\mathbf{x}}    \   {\mathcal{T}}_{\rm{s}}^z  
&= &        \int_{{\mathbf{x}}\in \rm{(interface)}}    d{\mathbf{x}}   \   \nabla \cdot {\mathbf{j}}_{\rm{s}}^{z}.          \\
&= &        \int      \    {\mathbf{j}}_{\rm{s}}^{z}     \cdot        d {\mathbf{S}}_{\rm{interface}}.
\label{eqn:e30}
 \end{eqnarray}

 In addition in our case, conduction electrons  cannot  enter the ferromagnet, which is an insulator. 
Then
the net spin current  pumped into the non-magnetic metal can be calculated
by integrating the SRT (eq. (\ref{eqn:e30}))
 over  the interface  (see Fig. \ref{fig:pumping}).
This is our spin pumping theory mediated by magnons
via the exchange interaction
at the interface between
a non-magnetic metal and  a ferromagnetic insulator.
The breaking of the spin conservation law for conduction electrons, $ {\mathcal{T}}_{\rm{s}}^z$,
 is  essential to our spin pumping theory.

\subsection{Quantum fluctuation}
\label{subsec:qf}

Our calculation
 (eqs. (\ref{eqn:e20})-(\ref{eqn:e21}))
gives, 
\begin{eqnarray}   
{\mathcal{T}}_{\rm{s}}^z  \stackrel{\Gamma \rightarrow 0}{\longrightarrow } 0.
\label{eqn:add1}
\end{eqnarray}   
Thus our formalism  (eq. (\ref{eqn:e30})) shows that
spin currents mediated by magnons cannot be pumped 
without quantum fluctuations;
this is the significant difference from the theory proposed by Tserkovnyak et al.\cite{mod2},
which is discussed at \S \ref{sec:Tserkovnyak} in detail. 
That is,  quantum fluctuations  are essential to spin pumping mediated by magnons 
as well as the exchange interaction between conduction electrons and ferromagnets (see eqs. (\ref{eqn:e20})-(\ref{eqn:e21}));
\begin{eqnarray}   
 {\mathcal{T}}_{\rm{s}}^z  \propto  J{\Gamma_0^2}.
 \label{eqn:add2}
 \end{eqnarray}   
This is the main feature of  our  quantum spin pumping theory  mediated by magnons.
In other words,
we have shown that
magnons accompanying the exchange interaction
cannot contribute to spin pumping 
without quantum fluctuations.

\begin{figure}[h]
\begin{center}
\includegraphics[width=12cm,clip]{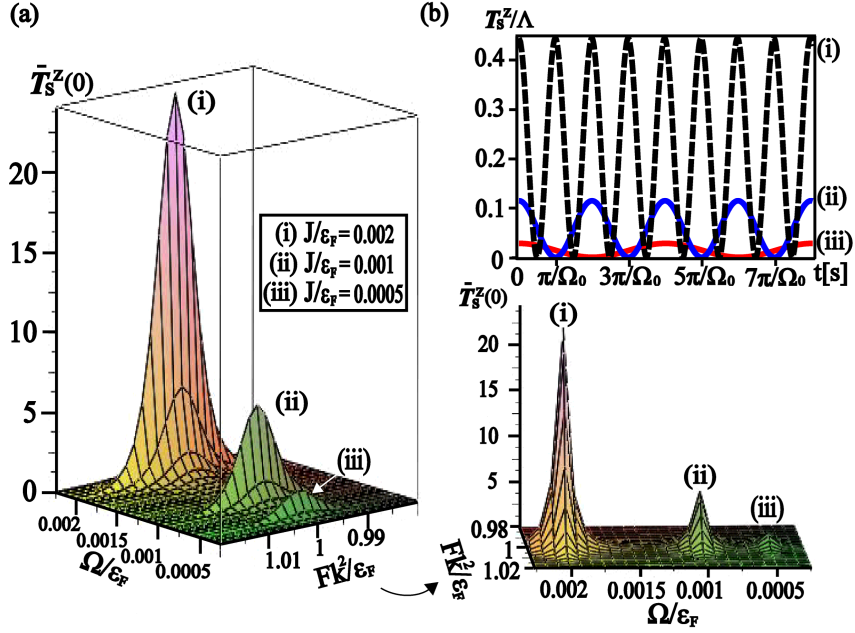}
\caption{(Color online)
(a)
The angular frequency dependence of the SRT; 
$\bar {\mathcal{T}}_{{\rm{s}}  }^{z}(0)$. 
A sharp peak exists on the point 
where the resonance condition, $\Omega = J$, is satisfied.
Each quantity, 
$\bar {\mathcal{T}}_{{\rm{s}}  }^{z}(\pm 1)$,
has the same structure with $\bar {\mathcal{T}}_{{\rm{s}}  }^{z}(0)$.
(b) The time evolution of the SRT, 
$ {\mathcal{T}}_{{\rm{s}}  }^{z}  /\Lambda   $,
at the resonance point ($\Omega =J$);
(i) $   0.446   \times   {\rm{cos}}^2 (\Omega_0 t) $,
(ii) $   0.112 \times    {\rm{cos}}^2 (\Omega_0 t/2)  $,
(iii) $   0.0281 \times    {\rm{cos}}^2 (\Omega_0 t/4)  $,
where $\Omega _0 \equiv  1.70 \times 10^{13}  $ $s^{-1}$.
 \label{fig:Fourier} }
\end{center}
\end{figure}

\subsection{Resonance}
\label{subsec:pump}

The SRT (eq. (\ref{eqn:e20})), $ {\mathcal{T}}_{{\rm{s}}  }^{z}$, is rewritten as (see also Appendix \ref{sec:SRT})
\begin{eqnarray}
            {\mathcal{T}}_{\rm{s}}^z    
             &\equiv &                                        \Lambda        \int_0^{\infty }    ( \sqrt{\frac{F}{\epsilon _{\rm{F}}}} dk) \    \bar {\mathcal{T}}_{{\rm{s}}  }^{z}, \ 
                                                                             \rm{where}               \label{eqn:e01}            \\                                 
 \bar {\mathcal{T}}_{{\rm{s}}  }^{z}
           &= &                                                  \sum_{n=0, \pm 1}  \bar {\mathcal{T}}_{\rm{s}}^{z} (n)   \   {\rm{e}}^{2in\Omega t},  \\
      \Lambda   &=&                                                                    \frac{\sqrt{\epsilon _{\rm{F}}} S  {{ {\Gamma} _0}}^2}{4(2\pi \sqrt{F})^3  }.  
\label{eqn:e101}
\end{eqnarray}
The variable, $\bar {\mathcal{T}}_{{\rm{s}}  }^{z} $, is the  dimensionless SRT as a function of 
the wavenumber for conduction electrons $ \sqrt{F/\epsilon _{\rm{F}}} k$, 
the angular frequency  of the applied  transverse magnetic field $\Omega /\epsilon _{\rm{F}}$,
the magnitude of the exchange interaction $J /\epsilon _{\rm{F}}$, and so on.
Fig. \ref{fig:Fourier} (a) represents the
\begin{equation}
\bar {\mathcal{T}}_{\rm{s}}^{z} (0), 
\end{equation}
which is the time average of $\bar {\mathcal{T}}_{\rm{s}}^{z}   $.
Fig. \ref{fig:Fourier} (b) shows  
\begin{equation}
 {\mathcal{T}}_{\rm{s}}^{z}   /\Lambda  
\end{equation}
(see eq. (\ref{eqn:e01})),
which describes the time evolution of the SRT.
 We  have set each parameter,   as a typical case,\cite{AndoPumping,xiao,spinwave,kittel}  as follows;
$ \epsilon _{\rm{F}} = 5.6  $ eV,
$ B/{\epsilon _{\rm{F}}} = 1.2\times 10^{-5} $,
$  T=300 $ K,
$ F=4$ eV {\AA}$^2$,
$ D=0.3$ eV {\AA}$^2$,
${\tau}=1$ ps   $={10^{-3}}\times {\tau}_{\rm{m}}$,
$a_0 =3$  {\AA},
$S=1/2$.

 Fig. \ref{fig:Fourier} (a)        shows that the SRT has a sharp peak as a result of the resonance
with the angular frequency, $\Omega $,  of the applied  transverse magnetic field. 
The sharp peak exists on the point 
where the condition, 
\begin{eqnarray}   
\Omega = J, 
\end{eqnarray}   
is satisfied.
This is because according to  eq. (\ref{eqn:e51}),
the localized spin acts as an effective magnetic field along the quantization axis $J$, 
which is far larger than the applied magnetic field $B$;
\begin{eqnarray}   
   {\cal{H}}_{\rm{ex}} ^{S=1/2}   =   -J  {\int_{{\mathbf{x}}\in \text{(interface)}}} d {\mathbf{x}}  {\ }   s^z ({\mathbf{x}},t).  
  \end{eqnarray}    
This fact (i.e. resonance condition) is useful to enhance the spin pumping effect 
because the angular frequency of 
a transverse magnetic field is under our control.
In addition Fig. \ref{fig:Fourier} (a) also shows that
the stronger the exchange interaction  becomes,
the larger  the SRT does.
 Fig. \ref{fig:Fourier} (b)        shows
$ {\mathcal{T}}_{{\rm{s}}  }^{z}  /\Lambda   $
at the resonance point ($\Omega =J$).
Each period reads 
\begin{equation}
\pi/\Omega.
\end{equation}

\section{Distinction from the  Theory Proposed by Tserkovnyak et al.}
\label{sec:Tserkovnyak}

Last, let us discuss  the distinction
between our quantum spin pumping theory\cite{nakatanote}
and the one proposed by Tserkovnyak et al.\cite{mod2} 
It should be noted that,
as has been mentioned in their article\cite{mod2}  
(see \S VIII. SUMMARY AND OUTLOOK in their article\cite{mod2}),
they have phenomenologically treated the spin-flip scattering processes,
which we have regarded as the most important processes for spin pumping. 
Nevertheless,
now their spin pumping theory has been widely used for interpreting  vast experimental results,\cite{AndoPumping,Kurebayashi,pumping5,pumping6}
in particular by experimentalists.
Thus, it would be significant to clarify the difference between our quantum spin pumping theory and the one proposed by Tserkovnyak et al.

\subsection{The spin pumping theory proposed by Tserkovnyak et al.}
\label{subsec:pump}

According to the phenomenological\cite{mod2,pumping5} spin pumping theory  by Tserkovnyak et al. and their notation,\cite{bauerrev,mod2}
the pumped spin current  ${\mathbf{I}}_{\rm{s{\mathchar`-}pump}}$ reads
\begin{eqnarray}
 {\mathbf{I}}_{\rm{s{\mathchar`-}pump}} = {\rm{G}}_{\perp }^{(\rm{R})} {\mathbf{m}}\times  \dot {\mathbf{m}}
                                                                               +    {\rm{G}}_{\perp }^{(\rm{I})}   \dot {\mathbf{m}},
\label{eqn:1}                                                                               
 \end{eqnarray}                                                                              
where the dot denotes the time derivative.
 We have taken ${e}=1$, and
 $ {\mathbf{m}}({\mathbf{x}},t)$  denotes a unit vector along the magnetization direction;
 they have treated  $ {\mathbf{m}}({\mathbf{x}},t)$ as  classical variables.
The variable ${\rm{G}}_{\perp }$ is the complex-valued mixing conductance that depends on the material;\cite{yunoki,conductance}
${\rm{G}}_{\perp } =  {\rm{G}}_{\perp }^{(\rm{R})} + i  {\rm{G}}_{\perp }^{(\rm{I})}$.

\subsection{Landau-Lifshitz-Gilbert equation}
\label{subsec:LLG}

The magnetization dynamics of ferromagnets can be described by the  Landau-Lifshitz-Gilbert (LLG) eq.;
\begin{eqnarray}
\dot {\mathbf{m}} = \gamma   {\mathbf{H}}_{\rm{eff}}  \times   {\mathbf{m}}  + \alpha   {\mathbf{m}} \times \dot {\mathbf{m}},
\label{eqn:2}
\end{eqnarray}
where $ \gamma $ is the gyro-magnetic ratio and
$\alpha $ is the Gilbert damping constant that determines the magnetization dissipation rate.
Here it should be emphasized that
though this Gilbert damping constant, $\alpha $,
was  phenomenologically introduced,\cite{Gilbert}
 it can be derived microscopically
 by considering a whole system including spin relaxation;\cite{tatara2}
 thus the effect of  the exchange coupling to conduction electrons should be considered to
 have already been included into this Gilbert damping term.

The effective magnetic field is set as  
 \begin{eqnarray}
 {\mathbf{H}}_{\rm{eff}}  = (\Gamma (t), 0, B),
  \end{eqnarray}
where $\Gamma (t)$  represents a time-dependent transverse magnetic field.
The  LLG   eq., eq. (\ref{eqn:2}),    becomes
\begin{equation}
\begin{pmatrix}
    \dot {\rm{m}}^x \\       \dot {\rm{m}}^y  \\     \dot {\rm{m}}^z   
\end{pmatrix}
= \gamma     
\begin{pmatrix}
  -B{\rm{m}^y}  \\    B{\rm{m}^x} -\Gamma {\rm{m}^z}    \\     \Gamma {\rm{m}^y} 
\end{pmatrix}
+\alpha   
\begin{pmatrix}
   {\rm{m}^y} \dot {{\rm{m}} }^z   -\dot {{\rm{m}} }^y  {\rm{m}^z}  \\ 
      {\rm{m}^z} \dot {{\rm{m}} }^x   -\dot {{\rm{m}} }^z  {\rm{m}^x}  \\     
  {\rm{m}^x} \dot {{\rm{m}} }^y   -\dot {{\rm{m}} }^x  {\rm{m}^y} 
\end{pmatrix}.
\label{eqn:2-2}
\end{equation}

\subsection{Pumped spin currents based on the theory by Tserkovnyak et al. at finite temperature}
\label{subsec:current}

Eq. (\ref{eqn:2-2})  is substituted into  ${\rm{I}}_{\rm{s{\mathchar`-}pump}}^z$, eq. (\ref{eqn:1});
we include the contribution of the Gilbert damping term, which depends on the materials,
up to   ${\cal{O}}(\alpha )$;
$ \alpha  \sim 10^{-3}, 10^{-2} $ for ${\rm{Ni}}_{81}{\rm{Fe}}_{19}$ (metal),\cite{AndoPumping} and 
$ \alpha  \sim 10^{-5} $ for ${\rm{Y}}_{3}{\rm{Fe}}_{5}{\rm{O}}_{12}$ (insulator),\cite{spinwave}
as examples.
Their theory is applicable to both ferromagnetic metals and insulators.\cite{battery}

Consequently, the z-component of the pumped spin current reads
\begin{eqnarray}
{\rm{I}}_{\rm{s{\mathchar`-}pump}}^z    
&=&          {\rm{G}}_{\perp }^{(\rm{R})}  \Big\{
   \gamma B  [         ({\rm{m}}^x)^2+ ({\rm{m}}^y)^2]   - \gamma \Gamma  {\rm{m}}^x  {\rm{m}}^z           
-   \alpha \gamma \Gamma   [{\rm{m}}^x  {\rm{m}}^y  {\rm{m}}^x + ({\rm{m}}^y)^3  + ({\rm{m}}^z)^2  {\rm{m}}^y] \Big\}       \nonumber    \\
&+&    {\rm{G}}_{\perp }^{(\rm{I})}  \Big\{      \alpha  \gamma    \{   B  [({\rm{m}}^x)^2+ ({\rm{m}}^y)^2]    -  \Gamma  {\rm{m}}^x   {\rm{m}}^z   \}     
+\gamma \Gamma {\rm{m}}^y \Big\}   + {\cal{O}}(\alpha ^2).               \\
  & \stackrel{\Gamma \rightarrow 0}{\longrightarrow }&
 [ {\rm{G}}_{\perp }^{(\rm{R})} +   \alpha  {\rm{G}}_{\perp }^{(\rm{I})}]
   \gamma B  [         ({\rm{m}}^x)^2+ ({\rm{m}}^y)^2].          \label{eqn:3}                \\
  & \stackrel{{\rm{G}}_{\perp }^{(\rm{I})}  \rightarrow 0}{\longrightarrow }&
  {\rm{G}}_{\perp }^{(\rm{R})}      \gamma B  [         ({\rm{m}}^x)^2+ ({\rm{m}}^y)^2].     
\label{eqn:3-2}                                   
\end{eqnarray}

At finite temperature,
the magnetization is thermally activated; $\dot {\mathbf{m}}\not=0$.\cite{xiao}
Then the time derivative of the z-component means
\begin{eqnarray}
 \dot {\rm{m}}^z 
 & =&  \gamma \Gamma {\rm{m}}^y + \alpha \gamma  \{B [({\rm{m}}^x)^2+ ({\rm{m}}^y)^2]   -\Gamma {\rm{m}}^x  {\rm{m}}^z\}  + {\cal{O}}(\alpha ^2). \label{eqn:e4-4}   \\
 &\stackrel{\Gamma \rightarrow 0}{\longrightarrow } &\alpha \gamma  B [({\rm{m}}^x)^2+ ({\rm{m}}^y)^2].  \label{eqn:e4}   \\
& \not=& 0.                                                                               
\label{eqn:4-3}                             
\end{eqnarray}

\subsection{Distinction}
\label{subsec:dis}

Eqs. (\ref{eqn:3}), (\ref{eqn:3-2})   and  (\ref{eqn:4-3}) mean that,
within the framework by  Tserkovnyak et al. with the LLG eq.,
they may  gain spin currents at finite temperature 
if only the magnetic field along the z-axis, $B$, is applied;
\begin{eqnarray}
 {\rm{I}}_{\rm{s{\mathchar`-}pump}}^z  \stackrel{\Gamma \rightarrow 0 (B\not=0)}{\nrightarrow }   0.                                                                                        
\label{eqn:add4}                             
\end{eqnarray}
That is, 
the spin pumping theory  by Tserkovnyak et al.\cite{mod2,bauerrev,battery}  with the LLG eq.   concludes that 
spin currents may be pumped at  finite temperature without time-dependent transverse magnetic fields.

On the other hand, as discussed in the last section (see eqs. (\ref{eqn:add1}) and  (\ref{eqn:add2})),
our approach based on the Schwinger-Keldysh formalism gives different result;
\begin{eqnarray}   
{\mathcal{T}}_{\rm{s}}^z  \stackrel{\Gamma \rightarrow 0}{\longrightarrow } 0.
\label{eqn:add3}
\end{eqnarray}   
That is, our quantum spin pumping theory means that
spin currents mediated by magnons cannot be pumped  without quantum fluctuations 
(i.e. time-dependent transverse magnetic fields, $\Gamma (t)$);
quantum fluctuations  are essential to spin pumping mediated by magnons 
as well as the exchange interaction between conduction electrons and ferromagnets.

This (eqs. (\ref{eqn:add3}) and (\ref{eqn:add4})) is the significant distinction 
between our quantum spin pumping theory and the one proposed by Tserkovnyak et al.

\section{Summary and Discussion}
\label{sec:sum}

We have  microscopically studied   quantum spin pumping  mediated by magnons
by evaluating the SRT at room temperature.
Localized spins of the ferromagnetic insulator 
lose spin angular momentum  by emitting a magnon
and conduction electrons flip  from down to up by absorbing the momentum.
Thus our formalism contains no phenomenological treatments of spin-flip scattering processes.
The SRT breaking the spin conservation law 
 represents the (net) spin current 
mediated  by magnons, which is pumped into the adjacent non-magnetic  metal.

Through the Schwinger-Keldysh formalism,
we have  clarified that 
quantum fluctuations (i.e. time-dependent transverse magnetic fields)
induce a net spin current, which can be enhanced through a resonance structure 
as a function of the angular frequency of the applied transverse field.
We conclude that the breaking of the spin conservation law and 
quantum fluctuations are essential to 
quantum spin pumping mediated by magnons
accompanying the exchange interaction.

In this paper,
we have theoretically introduced  and  defined the interface as an effective area where 
the Fermi gas (conduction electrons)  and 
the Bose gas (magnons) coexist to interact.
Though the behavior of the quantum spin pumping effect mediated by magnons
can be qualitatively captured by calculating the SRT,
the theoretical estimation for the volume, in particular the width, of the interface 
is  essential 
for the quantitative understanding.
Of course the width of the interface may be
roughly supposed to be
of the order of the lattice constant,
but
we consider  microscopic derivation 
 (so called proximity effects)
is an urgent theoretical issue.

In addition,
though  we have not executed vertex corrections at the present study
in order to clarify that  pumped spin currents are generated purely by quantum fluctuations,
calculating the SRT under the effects of vertex corrections (i.e. multiple scatterings of impurities) as well as quantum fluctuations is a significant theoretical issue.
Also from viewpoints of theoretical aspects (i.e. quantum field theory),
it might be better to execute vertex corrections.  
Thus, we would like to  tackle this issue in the near future.

In the present study,
we have focused exclusively on the dynamics at the interface ($J\not=0$), 
where  pumped spin currents are generated with
interchanging spin angular momentum  between conduction electrons and magnons.
Then, our quantum spin pumping theory microscopically well describes the dynamics of generating pumped spin currents,
in particular, the spin-flip scattering processes, which  we have regarded as the most important processes for spin pumping.
This is the strong point of our theory; note that
the theory proposed by Tserkovnyak et al. has phenomenologically treated spin-flip processes.\cite{mod2}
On the other hand, 
the dynamics of pumped spin currents in the non-magnetic metal ($J=0$), 
i.e. how the pumped spin currents flow in the non-magnetic metal,
is out of the application (purpose) of the present our theory. 
That is,
though our formalism has microscopically captured the spin-flip scattering processes at the interface beyond phenomenology,
it does not cover the dynamics of pumped spin currents;
the dynamics under the existence of pumped spin currents such as 
the effect of   localized spins and conduction electrons  on pumped spin currents,
which is often indicated as  the $\beta $ term,\cite{tatara,beta1,beta2}
is out of the application (purpose) of the present study.
Therefore,
we would like to brush up our quantum spin pumping theory
to cover the dynamics of pumped spin currents in the non-magnetic metal;
how the pumped spin currents flow in the non-magnetic metal.
We recognize that clarifying these issues is significant from viewpoints of applications
as well as fundamental science.

Last,
we have revealed that
magnons accompanying the exchange interaction
cannot contribute to quantum spin pumping  without quantum fluctuations.
Here it should be stressed that in our formalism,
the meditation of spin angular momentum is restricted to only magnons.
We will take the effect of phonons into account 
and develop our theory as a more rigorous formalism
to reveal the microscopic (quantum) dynamics of the magnon splitting.\cite{sandweg,Kurebayashi} 
Moreover,
we are also interested in the contribution of magnons to quantum spin pumping
under a spatially nonuniform magnetization.

\section*{Acknowledgements}
We would like to thank  K. Totsuka for stimulating the study,
and G. Tatara for reading the manuscript and useful comments.
We are also grateful to  T. Takahashi and Y. Korai  for fruitful discussion,
N. Sago  for helpful correspondence on numerical calculations,
and K. Ando for sending the invaluable presentation file, prior to the publication,
on the 6th International School and Conference on Spintronics and Quantum Information Technology (\textit{Spintech6}).
We also would like to thank   S. Onoda  and T. Oka  for crucial comments  at
the 26th Nishinomiya-Yukawa Memorial International Workshop 
\textit{Novel Quantum States in Condensed Matter} (\textit{NQS2011}).

We are supported by the Grant-in-Aid for the Global COE Program
"The Next Generation of Physics, Spun from Universality and Emergence"
from the Ministry of Education, Culture, Sports, Science and Technology (MEXT) of Japan.

\appendix 
\section{Langreth Method}
\label{sec:langreth}

In this section,
we show how to calculate  eqs. (\ref{eqn:e6})-(\ref{eqn:e9}) in detail.

First, we briefly show the Langreth method,\cite{nakatatatara,haug,tatara,new} 
which is useful to evaluate the perturbation expansion of  the Keldysh ( or contour-ordered) Green's function\cite{kamenev,kita,new,tatara,haug}
in subsection \ref{subsec:concrete}.
Second,
we introduce the point of  the bosonic Keldysh Green's functions\cite{kamenev,kita,new,rammer} in subsection \ref{subsec:boson}.
Last, the detailed calculations of eqs. (\ref{eqn:e6})-(\ref{eqn:e9}) are represented
in subsection \ref{subsec:ours}.

We omit the label, ${\mathbf{x}}$,
when it is not relevant.

\subsection{The Schwinger-Keldysh closed time path; a concrete example}
\label{subsec:concrete}

For simplicity here, we consider the perturbative term, $V^{\prime}$,
\begin{eqnarray}
V^{\prime} =  \Gamma (t) \int   d{\mathbf{x}}  [ a(\mathbf{x}, t)+a^{\dagger }(\mathbf{x}, t)],         
\label{eqn:appe2}
\end{eqnarray}
and evaluate the expectation value of the bosonic annihilate operator, $a$, as an example;
\begin{eqnarray}
\langle a(\tau) \rangle &=& \Big{\langle}\text{T}_{\text{c}} \ a(\tau ) \text{exp}[-i\int_{\text{c}} d\tau ' V(\tau ') ]  \Big{\rangle}   \\         
           & =&  -i\int d{\mathbf{x}}^{\prime} 
                 \int_{\text{c}} d\tau ' \Gamma (\tau ')  \langle \text{T}_{\text{c}} \ a(\mathbf{x}, \tau)a^{\dagger }(\mathbf{x'}, \tau ') \rangle    + {\cal{O}}(\Gamma ^2)\\
            & \equiv  &   -i \int d{\mathbf{x}}^{\prime}   {\cal{I}}.\;
\label{eqn:appe3}
\end{eqnarray}
Here $\text{T}_{\text{c}}$ is the path-ordering operator defined 
on   the  Schwinger-Keldysh closed time path,\cite{rammer}  $\text{c}$ (see Fig. A$\cdot $1 in our manuscript\cite{nakatatatara,kita,tatara,haug})
We express  the  Schwinger-Keldysh closed time path   as a sum of the forward path,
$c_{\rightarrow }$, and the backward path, $c_{\leftarrow }$;
$ c=c_{\rightarrow} + c_{\leftarrow} $.\cite{rammer,kita,tatara,haug}
We take $\tau $ which denotes the contour variable defined on the  Schwinger-Keldysh closed time path
on forward path, ${\rm{c}}_{\rightarrow }$.
Even when $\tau $ is located  on backward path, ${\rm{c}}_{\leftarrow }$, 
the result of this calculation is invariant
because each Green's function, $\rm{G}^{\rm{r}} $, $\rm{G}^{\rm{a}} $, $\rm{G}^{\rm{<}} $, $\rm{G}^{\rm{>}} $,
is not independent;\cite{kita,kamenev,new,tatara} they obey, 
\begin{eqnarray}
\rm{G}^{\rm{r}}  - \rm{G}^{\rm{a}}  = \rm{G}^{\rm{>}}  - \rm{G}^{\rm{<}}.
\end{eqnarray} 
Note that this relation comes into effect also for the fermionic case;\cite{kita,kamenev,new,tatara}
\begin{eqnarray}
 {\mathcal{G}}^{\text{r}}  - {\mathcal{G}}^{\text{a}}  = {\mathcal{G}}^{\text{>}}  - {\mathcal{G}}^{\text{<}}.
 \end{eqnarray}

The integral on  the  Schwinger-Keldysh closed time path  of eq. (\ref{eqn:appe3}), $  {\cal{I}} $,  is executed 
by taking an identity\cite{kita,new} into account 
\begin{equation}
\begin{split}
\int_{{\text{c}}} d\tau^{\prime} &= 
\int_{{\text{c}}_{\rightarrow }} d\tau^{\prime}
+ \int_{{\text{c}}_{\leftarrow  }} d\tau^{\prime}, \\
  \;
\end{split}
\label{eqn:kel}
\end{equation}
as (see also eq. (\ref{eqn:kel})  in Appendix \ref{subsec:boson})
\begin{equation}
\begin{split}
{\cal{I}} = i\int_{-\infty }^{\infty } dt^{\prime} \Gamma (t^{\prime}) [ {\rm{G}}^{\rm{t}}(t ,t^{\prime}) -  {\rm{G}}^{\rm{<}}(t ,t^{\prime})] . \;
\end{split}
\label{eqn:appe5}
\end{equation}
By using the relation,
$ {\rm{G}}^{\text{r}}(t,t') = {\rm{G}}^{\text{t}}(t,t') - {\rm{G}}^{\text{<}}(t,t') $,
we obtain 
\begin{equation}
\begin{split}
{\cal{I}} = i\int_{-\infty }^{\infty } dt^{\prime} \Gamma (t^{\prime})  {\rm{G}}^{\rm{r}}(t ,t^{\prime}). 
\end{split}
\label{eqn:appe5}
\end{equation}

\subsection{Bosonic Keldysh Green's function}
\label{subsec:boson}

In this subsection,
we show the point of  the bosonic Keldysh Green's functions.

The bosonic Keldysh Green's function, ${\rm{G}}(\tau,  \tau^{\prime}) $, is defined as\cite{kamenev,kita,new}
\begin{equation}
{\rm{G}}(\tau,  \tau^{\prime}) 
:= -i \langle  \text{T}_{\text{c}} a(\tau) a^{\dagger }( \tau^{\prime})   \rangle. 
\end{equation}
Depending on the points 
where $\tau$ and   $\tau^{\prime}$ are located on the  Schwinger-Keldysh closed time path 
(i.e. $\rm{c} =  {{\text{c}}_{\rightarrow  }}+ {{\text{c}}_{\leftarrow  }}$), 
the bosonic Keldysh Green's function is expressed as\cite{kamenev,kita,new}
\begin{equation}
{\rm{G}}(\tau,  \tau^{\prime}) 
=  \begin{cases}
             {\rm{G}}^{\rm{<}}(t, t^{\prime}) = -i \langle   a^{\dagger }(t^{\prime}) a(t)  \rangle, &  {\rm{when}}  \  
                                                                         \tau\in   {{\text{c}}_{\rightarrow  }},  \tau^{\prime} \in    {{\text{c}}_{\leftarrow  }}.     \\
            {\rm{G}}^{\rm{>}}(t, t^{\prime})=  -i \langle   a(t)  a^{\dagger }(t^{\prime})  \rangle, &  {\rm{when}}  \  
                                                                         \tau\in   {{\text{c}}_{\leftarrow  }},  \tau^{\prime} \in    {{\text{c}}_{\rightarrow  }}.  \\
           {\rm{G}}^{\rm{t}}(t, t^{\prime}),  &  {\rm{when}}    \  \tau\in   {{\text{c}}_{\rightarrow  }},  \tau^{\prime} \in    {{\text{c}}_{\rightarrow  }}.  \\
            {\rm{G}}^{\bar {\rm{t}}}(t, t^{\prime}),  & {\rm{when}}    \  \tau\in   {{\text{c}}_{\leftarrow  }},  \tau^{\prime} \in    {{\text{c}}_{\leftarrow  }}. 
            \label{eqn:kel}
      \end{cases}
 \end{equation}                                     

It should be noted that each Green's function is not independent;\cite{kamenev,kita,new} 
\begin{eqnarray}
\rm{G}^{\rm{r}}  - \rm{G}^{\rm{a}}  = \rm{G}^{\rm{>}}  - \rm{G}^{\rm{<}}
\Leftrightarrow  \rm{G}^{\rm{t}}  + \rm{G}^{\bar {\rm{t}}}  = \rm{G}^{\rm{>}}  + \rm{G}^{\rm{<}}.
\label{eqn:appe10}
\end{eqnarray} 
In addition, these relations\cite{new,rammer} would be useful on calculation;
\begin{eqnarray}
 \rm{G}^{\rm{t}} &=& \rm{G}^{\rm{a}} + \rm{G}^{\rm{>}},    \\ 
  \rm{G}^{\bar {\rm{t}}} &=& \rm{G}^{\rm{<}} -\rm{G}^{\rm{a}},  \\
  \rm{G}^{\rm{r}} &=&  \rm{G}^{\rm{t}} -  \rm{G}^{\rm{<}}           \\
                                &=&  \rm{G}^{\rm{>}}  - \rm{G}^{\bar {\rm{t}}},         \\
   \langle   a^{\dagger }(t) a(t)  \rangle  &=& i {\rm{G}}^{\rm{<}} (t, t).  
\end{eqnarray} 

By executing the Fourier transformation,
the lesser and greater Green's functions for free bosons  become\cite{kamenev,new}
\begin{eqnarray}
{\rm{G}}^{\rm{<}}_{   {\mathbf{k^{\prime}}}, \omega^{\prime} }  
&=& -f_{\rm{B}}(\omega^{\prime}) 
     ( {\rm{G}}^{\rm{a}}_{   {\mathbf{k^{\prime}}}, \omega^{\prime} }-  {\rm{G}}^{\rm{r}}_{   {\mathbf{k^{\prime}}}, \omega^{\prime} })       \\
 &=&-2 \pi   i  f_{\rm{B}}(\omega^{\prime}) \delta (\omega^{\prime}- \omega _{ {\mathbf{k^{\prime}}}}),                                                              \\
 {\rm{G}}^{\rm{>}}_{   {\mathbf{k^{\prime}}}, \omega^{\prime} }  
 &=&-2 \pi   i  [1+f_{\rm{B}}(\omega^{\prime})] \delta (\omega^{\prime}- \omega _{ {\mathbf{k^{\prime}}}}),   \\
   {\rm{G}}^{\rm{K}}_{   {\mathbf{k^{\prime}}}, \omega^{\prime} }
&:= &  {\rm{G}}^{\rm{<}}_{   {\mathbf{k^{\prime}}}, \omega^{\prime} }   +  {\rm{G}}^{\rm{>}}_{   {\mathbf{k^{\prime}}}, \omega^{\prime} }     \\
 &=&-2 \pi   i  [1+2f_{\rm{B}}(\omega^{\prime})] \delta (\omega^{\prime}- \omega _{ {\mathbf{k^{\prime}}}})                                                         \\
&=&2 i   {\rm{Im}}  {\rm{G}}^{\rm{r}}_{   {\mathbf{k^{\prime}}}, \omega^{\prime} }  {\rm{coth}}(\beta \omega^{\prime} /2).
\end{eqnarray}
The last one, $ {\rm{G}}^{\rm{K}}$, represents the Keldysh Green's function \cite{kamenev, kita}
and 
the relation is called the bosonic fluctuation-dissipation theorem.\cite{rammer}

\

\noindent{\textit{Fermionic Keldysh Green's function}}

It would be useful to compare with
the (spinless) Fermionic Keldysh Green's function, ${\mathcal{G}} (\tau,  \tau^{\prime}) $, 
which is defined as\cite{kamenev,kita,new,tatara,haug}
\begin{equation}
{\mathcal{G}}(\tau,  \tau^{\prime}) 
:= -i \langle  \text{T}_{\text{c}} c(\tau) c^{\dagger }( \tau^{\prime})   \rangle. 
\end{equation}
Depending on the points 
where $\tau$ and   $\tau^{\prime}$ are located on the  Schwinger-Keldysh closed time path 
(i.e. $\rm{c} =  {{\text{c}}_{\rightarrow  }}+ {{\text{c}}_{\leftarrow  }}$), 
the Fermionic Keldysh Green's function is expressed as\cite{kamenev,kita,new,tatara,haug}
\begin{equation}
{\mathcal{G}}(\tau,  \tau^{\prime}) 
=  \begin{cases}
             {\mathcal{G}}^{\rm{<}}(t, t^{\prime}) = i \langle   c^{\dagger }(t^{\prime}) c(t)  \rangle, &  {\rm{when}}  \  
                                                                         \tau\in   {{\text{c}}_{\rightarrow  }},  \tau^{\prime} \in    {{\text{c}}_{\leftarrow  }}.     \\
            {\mathcal{G}}^{\rm{>}}(t, t^{\prime})=  -i \langle   c(t)  c^{\dagger }(t^{\prime})  \rangle, &  {\rm{when}}  \  
                                                                         \tau\in   {{\text{c}}_{\leftarrow  }},  \tau^{\prime} \in    {{\text{c}}_{\rightarrow  }}.  \\
           {\mathcal{G}}^{\rm{t}}(t, t^{\prime}),  &  {\rm{when}}    \  \tau\in   {{\text{c}}_{\rightarrow  }},  \tau^{\prime} \in    {{\text{c}}_{\rightarrow  }}.  \\
            {\mathcal{G}}^{\bar {\rm{t}}}(t, t^{\prime}),  & {\rm{when}}    \  \tau\in   {{\text{c}}_{\leftarrow  }},  \tau^{\prime} \in    {{\text{c}}_{\leftarrow  }}. 
            \label{eqn:kelfermion}
      \end{cases}
 \end{equation}                                     
Note that with reflecting the statistical properties,
the sign of the lesser Green's function is opposite from the bosonic case.
In addition, they satisfy the relation;\cite{kamenev,kita,new,tatara,haug}
\begin{eqnarray}
\mathcal{G}^{\rm{r}}  - \mathcal{G}^{\rm{a}}  = \mathcal{G}^{\rm{>}}  - \mathcal{G}^{\rm{<}}
\Leftrightarrow 
\mathcal{G}^{\rm{t}}  + \mathcal{G}^{\bar {\rm{t}}}  = \mathcal{G}^{\rm{>}}  + \mathcal{G}^{\rm{<}},
\label{eqn:appe10fermion}
\end{eqnarray} 
\begin{eqnarray}
\mathcal{G}^{\rm{t}}  &=& \mathcal{G}^{\rm{r}}  + \mathcal{G}^{\rm{<}}    \\
                                        &=& \mathcal{G}^{\rm{a}}  + \mathcal{G}^{\rm{>}},    \\
 \mathcal{G}^{\bar {\rm{t}}}  &=& -\mathcal{G}^{\rm{r}}  + \mathcal{G}^{\rm{>}}    \\
                                                    &=& -\mathcal{G}^{\rm{a}}  + \mathcal{G}^{\rm{<}}.
\label{eqn:appe101fermion}
\end{eqnarray} 

By executing the Fourier transformation,
the lesser and greater Green's functions for free Fermions  become\cite{kamenev,new,tatara,haug}
\begin{eqnarray}
{\mathcal{G}}^{\rm{<}}_{   {\mathbf{k}}, \omega }  
&= &f_{\rm{F}}(\omega) 
     ( {\mathcal{G}}^{\rm{a}}_{   {\mathbf{k}}, \omega }-  {\mathcal{G}}^{\rm{r}}_{   {\mathbf{k}}, \omega })       \\
 &=&2 \pi   i  f_{\rm{F}}(\omega) \delta (\omega- \omega _{ {\mathbf{k}}}),                                                              \\
 {\mathcal{G}}^{\rm{>}}_{   {\mathbf{k}}, \omega }  
 &=&-2 \pi   i  [1-f_{\rm{F}}(\omega)] \delta (\omega- \omega _{ {\mathbf{k}}}),   \\
   {\mathcal{G}}^{\rm{K}}_{   {\mathbf{k}}, \omega }
&:= &  {\mathcal{G}}^{\rm{<}}_{   {\mathbf{k}}, \omega }   +  {\mathcal{G}}^{\rm{>}}_{   {\mathbf{k}}, \omega }     \\
 &=&-2 \pi   i  [1-2f_{\rm{F}}(\omega )] \delta (\omega - \omega _{ {\mathbf{k}}})                                                         \\
&=&2 i   {\rm{Im}}  {\mathcal{G}}^{\rm{r}}_{   {\mathbf{k}}, \omega }  {\rm{tanh}}(\beta \omega /2).
\end{eqnarray}
The last one, $ {\mathcal{G}}^{\rm{K}}$, represents the Keldysh Green's function \cite{kamenev, kita}
and 
the relation is called the fermionic fluctuation-dissipation theorem.\cite{rammer,haug}

\subsection{Detail of the calculation; eqs. (\ref{eqn:e6})-(\ref{eqn:e9})}
\label{subsec:ours}

In this subsection,
we  will briefly show how to calculate eqs. (\ref{eqn:e6})-(\ref{eqn:e9}).

By  adopting  the Wick's theorem,\cite{kita,rammer}
the left-hand side (LHS) of  eq. (\ref{eqn:e6}) reads 
\begin{equation}
\begin{split}
({\rm{The\ LHS\ of\ }} {\rm{eq}}. (\ref{eqn:e6})) &= \frac{iJa_0^3  {\tilde S}}{2} 
                                                      \int d{\mathbf{x}}^{\prime}  \int d{\mathbf{x}}^{\prime \prime}
                                                      \int_{{\text{c}}} d\tau^{\prime}       \int_{{\text{c}}} d\tau^{\prime \prime}
                                                      \Gamma (\tau^{\prime})      \Gamma (\tau^{\prime \prime})    
                                                          \Big(1-\frac{ \langle \text{T}_{\text{c}} \  a^{\dagger }({\mathbf{x}}^{\prime}, \tau^{\prime}) 
                                                                                                                      a({\mathbf{x}}^{\prime }, \tau^{\prime})\rangle }{\tilde S} \Big)                                     \\     
                                              &\times      \langle \text{T}_{\text{c}} \ a({\mathbf{x}}^{\prime \prime}, \tau^{\prime \prime})
                                                                                                       a^{\dagger }({\mathbf{x}}, \tau) \rangle    
                                                                   \langle \text{T}_{\text{c}} \ c_{\downarrow }({\mathbf{x}}, \tau)
                                                                                                       c_{\downarrow }^{\dagger }({\mathbf{x}^{\prime}}, \tau^{\prime}) \rangle    
                                                                   \langle \text{T}_{\text{c}} \ c_{\uparrow }({\mathbf{x}}^{\prime}, \tau^{\prime})   c_{\uparrow }^{\dagger }({\mathbf{x}}, \tau ) \rangle. 
\label{eqn:a10}                                                                                                                                      
\end{split}
\end{equation}
By employing the relation;
\begin{eqnarray}
    \int_{{\text{c}}} d\tau^{\prime}       \int_{{\text{c}}} d\tau^{\prime \prime} 
   &=& \Big(   \int_{{\text{c}}_{\rightarrow }} d\tau^{\prime}
                 + \int_{{\text{c}}_{\leftarrow  }} d\tau^{\prime}   \Big)
                   \Big(   \int_{{\text{c}}_{\rightarrow }} d\tau^{\prime  \prime}
                   + \int_{{\text{c}}_{\leftarrow  }} d\tau^{\prime  \prime}   \Big)             \\
     &= &      \int_{{\text{c}}_{\rightarrow }} d\tau ^{\prime}         \int_{{\text{c}}_{\rightarrow }} d\tau^{\prime  \prime}
             +\int_{{\text{c}}_{\rightarrow }} d\tau^{\prime}   \int_{{\text{c}}_{\leftarrow  }} d\tau ^{\prime   \prime}    \nonumber  \\
&+&      \int_{{\text{c}}_{\leftarrow  }} d\tau^{\prime} \int_{{\text{c}}_{\rightarrow }} d\tau^{\prime   \prime}
          +   \int_{{\text{c}}_{\leftarrow  }} d\tau^{\prime} \int_{{\text{c}}_{\leftarrow  }} d\tau^{\prime  \prime},
\end{eqnarray}
and the Langreth method\cite{haug,tatara,new} with eqs. (\ref{eqn:kel}) and (\ref{eqn:kelfermion}),
the right-hand side (RHS) of eq. (\ref{eqn:a10}) can be expressed as 
\begin{equation}
\begin{split}
(\rm{The \  RHS \  of  \  } {\rm{eq}}. (\ref{eqn:a10})) &= \frac{Ja_0^3\tilde S}{2}  \int d{\mathbf{x}}^{\prime}  \int d{\mathbf{x}}^{\prime \prime}
                                                       \int_{-\infty }^{\infty }  dt^{\prime}  \int_{-\infty }^{\infty }  dt^{\prime  \prime}  \Gamma (t^{\prime})      \Gamma (t^{\prime \prime})  
                                            \Big(    1-\frac{i}{\tilde S}  {\rm{G}}^{\rm{<}}(t^{\prime}, t^{\prime})  \Big)             \\
                                                 &\times [{\rm{G}}^{\rm{t}}(t^{\prime \prime}, t) - {\rm{G}}^{\rm{>}}(t^{\prime \prime}, t) ]
                                                                   [{\cal{G}}_{\downarrow }^{\rm{t}}(t, t^{\prime}) {\cal{G}}_{\uparrow }^{\rm{t}}(t^{\prime}, t)      
                                                                    -   {\cal{G}}_{\downarrow }^{\rm{<}}(t, t^{\prime}) {\cal{G}}_{\uparrow }^{\rm{>}}(t^{\prime}, t)].
   \label{eqn:a11}                                                                    
\end{split}
\end{equation}
We also here have taken $\tau $  on forward path, ${\rm{c}}_{\rightarrow }$.
As discussed in the last subsection,
even when $\tau $ is located  on backward path, ${\rm{c}}_{\leftarrow }$, 
the result of this calculation is invariant.

Here it should be noted that 
\begin{equation}
{\rm{G}}^{\rm{t}}(t^{\prime \prime}, t) - {\rm{G}}^{\rm{>}}(t^{\prime \prime}, t) = {\rm{G}}^{\rm{a}}(t^{\prime \prime}, t). 
\end{equation}
Then, eq. (\ref{eqn:a11}) can be rewritten as 
\begin{equation}
\begin{split}
(\rm{The \  RHS \  of  \  } {\rm{eq}}. (\ref{eqn:a11})) &= \frac{Ja_0^3\tilde S}{2}  \int d{\mathbf{x}}^{\prime}  \int d{\mathbf{x}}^{\prime \prime}
                                                       \int_{-\infty }^{\infty }  dt^{\prime}  \int_{-\infty }^{\infty }  dt^{\prime  \prime}  \Gamma (t^{\prime})      \Gamma (t^{\prime \prime})  
                                                              \Big(    1-\frac{i}{\tilde S}  {\rm{G}}^{\rm{<}}(t^{\prime}, t^{\prime})  \Big)                 \\
                                                 &\times {\rm{G}}^{\rm{a}}(t^{\prime \prime}, t)
                                                                   [{\cal{G}}_{\downarrow }^{\rm{t}}(t, t^{\prime}) {\cal{G}}_{\uparrow }^{\rm{t}}(t^{\prime}, t)      
                                                                    -   {\cal{G}}_{\downarrow }^{\rm{<}}(t, t^{\prime}) {\cal{G}}_{\uparrow }^{\rm{>}}(t^{\prime}, t)].
   \label{eqn:a12}
\end{split}
\end{equation}
By executing Fourier transformation,
 we obtain the RHS of  eq. (\ref{eqn:e6}).

Through the same procedure,
remained terms are evaluated as follows;
\begin{equation}
\begin{split}
({\rm{The\ LHS\ of\ }}{\rm{eq}}. (\ref{eqn:e7})) &=  -  \frac{iJa_0^3  {\tilde S}}{2} 
                                                      \int d{\mathbf{x}}^{\prime}  \int d{\mathbf{x}}^{\prime \prime}
                                                      \int_{{\text{c}}} d\tau^{\prime}       \int_{{\text{c}}} d\tau^{\prime \prime}
                                                      \Gamma (\tau^{\prime})      \Gamma (\tau^{\prime \prime})    
                                                          \Big(1-\frac{ \langle \text{T}_{\text{c}} \  a^{\dagger }({\mathbf{x}}^{\prime}, \tau^{\prime}) 
                                                                                                                      a({\mathbf{x}}^{\prime }, \tau^{\prime })\rangle }{\tilde S} \Big)                                     \\     
                                              &\times      \langle \text{T}_{\text{c}} \ a({\mathbf{x}}, \tau)
                                                                                                       a^{\dagger }({\mathbf{x}}^{\prime  \prime}, \tau^{\prime  \prime}) \rangle    
                                                                   \langle \text{T}_{\text{c}} \ c_{\uparrow }({\mathbf{x}}, \tau)
                                                                                                       c_{\uparrow }^{\dagger }({\mathbf{x}^{\prime}}, \tau^{\prime}) \rangle    
                                                                   \langle \text{T}_{\text{c}} \ c_{\downarrow }({\mathbf{x}}^{\prime}, \tau^{\prime})   c_{\downarrow }^{\dagger }({\mathbf{x}}, \tau ) \rangle   \\
&= -\frac{Ja_0^3\tilde S}{2}  \int d{\mathbf{x}}^{\prime}  \int d{\mathbf{x}}^{\prime \prime}
                                                       \int_{-\infty }^{\infty }  dt^{\prime}  \int_{-\infty }^{\infty }  dt^{\prime  \prime}  \Gamma (t^{\prime})      \Gamma (t^{\prime \prime})  
                                                              \Big(    1-\frac{i}{\tilde S}  {\rm{G}}^{\rm{<}}(t^{\prime}, t^{\prime})  \Big)                 \\
                                                 &\times {\rm{G}}^{\rm{r}}(t, t^{\prime \prime})
                                                                   [{\cal{G}}_{\uparrow }^{\rm{t}}(t, t^{\prime}) {\cal{G}}_{\downarrow }^{\rm{t}}(t^{\prime}, t)      
                                                                    -   {\cal{G}}_{\uparrow }^{\rm{<}}(t, t^{\prime}) {\cal{G}}_{\downarrow }^{\rm{>}}(t^{\prime}, t)].
 \label{eqn:aaa}                                                                   
\end{split}
\end{equation}
\begin{equation}
\begin{split}
({\rm{The\ LHS\ of\ }}{\rm{eq}}. (\ref{eqn:e8})) &=    \frac{iJa_0^3  {\tilde S}}{2} 
                                                      \int d{\mathbf{x}}^{\prime}  \int d{\mathbf{x}}^{\prime \prime}
                                                      \int_{{\text{c}}} d\tau^{\prime}       \int_{{\text{c}}} d\tau^{\prime \prime}
                                                      \Gamma (\tau)      \Gamma (\tau^{\prime \prime})    
                                                          \Big(1-\frac{ \langle \text{T}_{\text{c}} \  a^{\dagger }({\mathbf{x}}^{\prime}, \tau^{\prime}) 
                                                                                                                      a({\mathbf{x}}^{\prime }, \tau^{\prime })\rangle }{\tilde S} \Big)                                     \\     
                                              &\times      \langle \text{T}_{\text{c}} \ a({\mathbf{x}}^{\prime}, \tau^{\prime})
                                                                                                       a^{\dagger }({\mathbf{x}}^{\prime  \prime}, \tau^{\prime  \prime}) \rangle    
                                                                   \langle \text{T}_{\text{c}} \ c_{\downarrow }({\mathbf{x}}, \tau)
                                                                                                       c_{\downarrow }^{\dagger }({\mathbf{x}^{\prime}}, \tau^{\prime}) \rangle    
                                                                   \langle \text{T}_{\text{c}} \ c_{\uparrow }({\mathbf{x}}^{\prime}, \tau^{\prime})   c_{\uparrow }^{\dagger }({\mathbf{x}}, \tau ) \rangle   \\
&= \frac{Ja_0^3\tilde S}{2}  \int d{\mathbf{x}}^{\prime}  \int d{\mathbf{x}}^{\prime \prime}
                                                       \int_{-\infty }^{\infty }  dt^{\prime}  \int_{-\infty }^{\infty }  dt^{\prime  \prime}  \Gamma (t)      \Gamma (t^{\prime \prime})  
                                                              \Big(    1-\frac{i}{\tilde S}  {\rm{G}}^{\rm{<}}(t^{\prime}, t^{\prime})  \Big)                 \\
                                                 &\times {\rm{G}}^{\rm{r}}(t^{\prime}, t^{\prime \prime})
                                                                   [{\cal{G}}_{\downarrow }^{\rm{t}}(t, t^{\prime}) {\cal{G}}_{\uparrow }^{\rm{t}}(t^{\prime}, t)      
                                                                    -   {\cal{G}}_{\downarrow }^{\rm{<}}(t, t^{\prime}) {\cal{G}}_{\uparrow }^{\rm{>}}(t^{\prime}, t)].
 \label{eqn:aaa2}                                                                      
\end{split}
\end{equation}
Note that we have adopted the relation;
$ {\rm{G}}^{\rm{t}} - {\rm{G}}^{\rm{<}}= {\rm{G}}^{\rm{r}} = {\rm{G}}^{\rm{>}} - {\rm{G}}^{\bar {\rm{t}}} $.
\begin{equation}
\begin{split}
({\rm{The\ LHS\ of\ }}{\rm{eq}}. (\ref{eqn:e9})) &=   - \frac{iJa_0^3  {\tilde S}}{2} 
                                                      \int d{\mathbf{x}}^{\prime}  \int d{\mathbf{x}}^{\prime \prime}
                                                      \int_{{\text{c}}} d\tau^{\prime}       \int_{{\text{c}}} d\tau^{\prime \prime}
                                                      \Gamma (\tau)      \Gamma (\tau^{\prime \prime})    
                                                          \Big(1-\frac{ \langle \text{T}_{\text{c}} \  a^{\dagger }({\mathbf{x}}^{\prime}, \tau^{\prime}) 
                                                                                                                      a({\mathbf{x}}^{\prime }, \tau^{\prime })\rangle }{\tilde S} \Big)                                     \\     
                                              &\times      \langle \text{T}_{\text{c}} \ a({\mathbf{x}}^{\prime \prime}, \tau^{\prime \prime})
                                                                                                       a^{\dagger }({\mathbf{x}}^{\prime  }, \tau^{\prime }) \rangle    
                                                                   \langle \text{T}_{\text{c}} \ c_{\uparrow }({\mathbf{x}}, \tau)
                                                                                                       c_{\uparrow }^{\dagger }({\mathbf{x}^{\prime}}, \tau^{\prime}) \rangle    
                                                                   \langle \text{T}_{\text{c}} \ c_{\downarrow }({\mathbf{x}}^{\prime}, \tau^{\prime})   c_{\downarrow }^{\dagger }({\mathbf{x}}, \tau ) \rangle   \\
&= -\frac{Ja_0^3\tilde S}{2}  \int d{\mathbf{x}}^{\prime}  \int d{\mathbf{x}}^{\prime \prime}
                                                       \int_{-\infty }^{\infty }  dt^{\prime}  \int_{-\infty }^{\infty }  dt^{\prime  \prime}  \Gamma (t)      \Gamma (t^{\prime \prime})  
                                                              \Big(    1-\frac{i}{\tilde S}  {\rm{G}}^{\rm{<}}(t^{\prime}, t^{\prime})  \Big)                 \\
                                                 &\times {\rm{G}}^{\rm{a}}(t^{\prime \prime}, t^{\prime})
                                                                   [{\cal{G}}_{\uparrow }^{\rm{t}}(t, t^{\prime}) {\cal{G}}_{\downarrow }^{\rm{t}}(t^{\prime}, t)      
                                                                    -   {\cal{G}}_{\uparrow }^{\rm{<}}(t, t^{\prime}) {\cal{G}}_{\downarrow }^{\rm{>}}(t^{\prime}, t)].
  \label{eqn:aaa3}                                                                     
\end{split}
\end{equation}
Note that we have employed the relation;
${\rm{G}}^{\rm{<}}-  {\rm{G}}^{\bar {\rm{t}}} ={\rm{G}}^{\rm{a}} ={\rm{G}}^{\rm{t}}-  {\rm{G}}^{ {\rm{>}}}$.

By using Fourier transformation,  
we obtain  the RHS of eqs. (\ref{eqn:e7})-(\ref{eqn:e9}).

\appendix 
\section{The Form of the Spin Relaxation Torque}
\label{sec:SRT}

Finally,
let us write down the explicit form of the SRT.
We have taken $\hbar =1$ and have introduced the dimensionless variables as follows;
$\bar {\cal{O}} \equiv {\cal{O}}/{\epsilon _{\rm{F}}}$, except
$ \bar T\equiv k_{\rm{B}}T/{\epsilon _{\rm{F}}}$ and  $\bar {\tau}\equiv {\epsilon _{\rm{F}}}\tau/(\hbar )$.
The dimensionless wavenumber for conduction electrons has been rewritten;
$ \sqrt{F/\epsilon _{\rm{F}}} k \equiv y$.

The SRT (i.e.  eq. (\ref{eqn:e01})),    $  {\mathcal{T}}_{\rm{s}}^z$,   reads
\begin{equation}
 {\cal{T}}_{\rm{s}}^z  = ({\rm{I}}) + ({\rm{II}}),
\end{equation}
where
\begin{eqnarray}
({\rm{I}}) &:=&  \bigg{\langle}    iJa_0^3  \sqrt{\frac{\tilde S}{2}}   a^{\dagger }({\mathbf{x}},t) 
                                                    \Big[1-\frac{ a^{\dagger }({\mathbf{x}},t) a({\mathbf{x}},t)}{4 \tilde S }\Big]
                                                      c^{\dagger }  ({\mathbf{x}},t) \sigma^{+}   c({\mathbf{x}},t )  \bigg{\rangle}                                                                    
  +  \bigg{\langle}   \frac{\Gamma (t)  c^{\dagger }  ({\mathbf{x}},t) \sigma^{+}   c({\mathbf{x}},t )  }{4i}  \bigg{\rangle}        \\                                        
 &=& \frac{ JS }{ 4 \pi }(\frac{\Gamma _0}{2})^2  \int \frac{d{ {\mathbf{k}}}}{(2\pi )^3} \int d\omega    
                                \Big[1-\frac{i}{\tilde S }     \int \frac{d{ {\mathbf{k}}}^{\prime}}{(2\pi )^3}  \int \frac{d{\omega}^{\prime} }{2\pi} 
                                   {\rm{G}}^{\rm{<}}_{{ {\mathbf{k}}}^{\prime}, \omega ^{\prime}}      \Big]        
                          \nonumber      \\
 &\times& \Bigg{\{}  
    \Big[   {\rm{e}}^{2i\Omega t} {\rm{G}}^{\rm{a}}_{0, \Omega } + {\rm{G}}^{\rm{a}}_{0, -\Omega }                                                      
               + ({\rm{e}}^{2i\Omega t}+1) {\rm{G}}^{\rm{r}}_{0, -\Omega }   
    \Big]                                                                                                                                                                                                                       
     \cdot     \Big[{\mathcal{G}}^{\rm{t}}_{\downarrow , {\mathbf{k}}, \omega -\Omega } 
              {\mathcal{G}}^{\rm{t}}_{\uparrow  , {\mathbf{k}}, \omega  }  
                -  {\mathcal{G}}^{\rm{<}}_{\downarrow , {\mathbf{k}}, \omega -\Omega } 
              {\mathcal{G}}^{\rm{>}}_{\uparrow  , {\mathbf{k}}, \omega  } 
    \Big]                                                                                                        \nonumber       \\
  &+& \Big[   {\rm{e}}^{-2i\Omega t} {\rm{G}}^{\rm{a}}_{0, -\Omega } + {\rm{G}}^{\rm{a}}_{0, \Omega }    
               + ({\rm{e}}^{-2i\Omega t}+1) {\rm{G}}^{\rm{r}}_{0, \Omega }   
    \Big]   
     \cdot \Big[{\mathcal{G}}^{\rm{t}}_{\downarrow , {\mathbf{k}}, \omega +\Omega } 
     {\mathcal{G}}^{\rm{t}}_{\uparrow  , {\mathbf{k}}, \omega  }   
                 -  {\mathcal{G}}^{\rm{<}}_{\downarrow , {\mathbf{k}}, \omega +\Omega } 
              {\mathcal{G}}^{\rm{>}}_{\uparrow  , {\mathbf{k}}, \omega  } 
    \Big]                                            
                         \Bigg{\}}    \nonumber         \\
    &+& {\cal{O}}(J^2)      + {\cal{O}}({(\Gamma_0)}^0)               \\
   &=&  {\bar{J}} \int  d\bar {\omega }       \int_0^{\infty } dy y^2  
        \Bigg[
                   \frac{  {\rm{e}}^{2i\Omega t} (\bar {\omega  }  -\bar{B}+\frac{i}{2\bar{\tau}_{\rm{m}}})  }{  (\bar {\Omega  }  -\bar{B})^2+ (\frac{1}{2\bar{\tau}_{\rm{m}}})^2  }
                   -  \frac{  ({\rm{e}}^{2i\Omega t} +2) (\bar {\Omega  }  +\bar{B}) + {\rm{e}}^{2i\Omega t}  
                      \frac{i}{2\bar{\tau}_{\rm{m}}})  }{  (\bar {\Omega  }  +\bar{B})^2+ (\frac{1}{2\bar{\tau}_{\rm{m}}})^2  }
        \Bigg]       \nonumber         \\    
& \times &  \Big[1-     \frac{i}{\tilde S } \int \frac{d{ {\mathbf{k}}}^{\prime}}{(2\pi )^3}    \int \frac{d{\omega}^{\prime} }{2\pi} 
                                 f_{\rm{B}} (\omega^{\prime})   \frac{-i/{\tau_{\rm{m}}}}{(\omega ^{\prime}-\omega _{{\mathbf{k^{\prime}}}})^2+[1/(2\tau_{\rm{m}})]^2}
\Big]         \nonumber       \\ 
     &\times &   \Bigg\{ \frac{1}{\Big[  \bar{\omega }- \bar{\Omega  } -y^2- (\bar{J}S+\frac{\bar{B}}{2}+\bar {\mu})  \Big]^2 + (\frac{1}{2\bar{\tau}})^2 }  \nonumber     \\
        &\times &                             \Bigg[
                                     \frac{i}{\bar{\tau}}  \frac{1}{{\rm{e}}^{[{{y}}^2 +(\bar{J}S+\bar{B}/2) -\bar{\mu} ]/{\bar{T}}}+1}   
                                                  +  \bar{\omega }- \bar{\Omega } -y^2- (\bar{J}S+\frac{\bar{B}}{2}+\bar {\mu})    -  \frac{i}{2\bar{\tau}}
                                    \Bigg]  \nonumber       \\
          & \times &  \frac{1}{\Big[  \bar{\omega } -y^2+ (\bar{J}S+\frac{\bar{B}}{2}+\bar {\mu})  \Big]^2 + (\frac{1}{2\bar{\tau}})^2 } 
                                    \Bigg[
                                     \frac{i}{\bar{\tau}}  \frac{1}{{\rm{e}}^{[{{y}}^2 -(\bar{J}S+\bar{B}/2) -\bar{\mu} ]/{\bar{T}}}+1}   
                                                  +  \bar{\omega } -y^2+ (\bar{J}S+\frac{\bar{B}}{2}+\bar {\mu})    -  \frac{i}{2\bar{\tau}}     \Bigg]  \nonumber        \\ 
       &-&     \frac{1}{{\bar{\tau}}^2}    \frac{1}{\Big[  \bar{\omega }- \bar{\Omega } -y^2- (\bar{J}S+\frac{\bar{B}}{2}+\bar {\mu})  \Big]^2 + (\frac{1}{2\bar{\tau}})^2 } 
                                  \cdot     \frac{1}{{\rm{e}}^{[{{y}}^2 +(\bar{J}S+\bar{B}/2) -\bar{\mu} ]/{\bar{T}}}+1}   \nonumber     \\
      & \times &       \frac{1}{\Big[  \bar{\omega } -y^2+ (\bar{J}S+\frac{\bar{B}}{2}+\bar {\mu})  \Big]^2 + (\frac{1}{2\bar{\tau}})^2 }   
                                     \cdot   \Big(   1-  \frac{1}{{\rm{e}}^{[{{y}}^2 -(\bar{J}S+\bar{B}/2) -\bar{\mu} ]/{\bar{T}}}+1 }  \Big)          
      \Bigg\}  \nonumber       \\
 &+& {\bar{J}} \int  d\bar {\omega }  \int_0^{\infty } dy y^2  
        \Bigg[
                   \frac{  {\rm{e}}^{-2i\Omega t} (\bar {\omega  }  -\bar{B}+\frac{i}{2\bar{\tau}_{\rm{m}}})  }{  (-\bar {\Omega  }  -\bar{B})^2+ (\frac{1}{2\bar{\tau}_{\rm{m}}})^2  }
                   -  \frac{  ({\rm{e}}^{-2i\Omega t} +2) (-\bar {\Omega  }  +\bar{B}) + {\rm{e}}^{-2i\Omega t}  
                      \frac{i}{2\bar{\tau}_{\rm{m}}})  }{  (-\bar {\Omega  }  +\bar{B})^2+ (\frac{1}{2\bar{\tau}_{\rm{m}}})^2  }
        \Bigg]        \nonumber        \\
&\times &  \Big[1-\frac{i}{\tilde S }  \int \frac{d{ {\mathbf{k}}}^{\prime}}{(2\pi )^3}   \int \frac{d{\omega}^{\prime} }{2\pi} 
                           f_{\rm{B}} (\omega^{\prime})   \frac{-i/{\tau_{\rm{m}}}}{(\omega ^{\prime}-\omega _{{\mathbf{k^{\prime}}}})^2+[1/(2\tau_{\rm{m}})]^2}    
 \Big]           \nonumber      \\ 
   &   \times &  \Bigg\{ \frac{1}{\Big[  \bar{\omega }+ \bar{\Omega } -y^2- (\bar{J}S+\frac{\bar{B}}{2}+\bar {\mu})  \Big]^2 + (\frac{1}{2\bar{\tau}})^2 }  \nonumber     \\
      &\times &                               \Bigg[
                                     \frac{i}{\bar{\tau}}  \frac{1}{{\rm{e}}^{[{{y}}^2 +(\bar{J}S+\bar{B}/2) -\bar{\mu} ]/{\bar{T}}}+1}   
                                                  +  \bar{\omega }+ \bar{\Omega } -y^2- (\bar{J}S+\frac{\bar{B}}{2}+\bar {\mu})    -  \frac{i}{2\bar{\tau}}
                                    \Bigg]   \nonumber         \\
      &    \times & \frac{1}{\Big[  \bar{\omega } -y^2+ (\bar{J}S+\frac{\bar{B}}{2}+\bar {\mu})  \Big]^2 + (\frac{1}{2\bar{\tau}})^2 } 
                                    \Bigg[
                                     \frac{i}{\bar{\tau}}  \frac{1}{{\rm{e}}^{[{{y}}^2 -(\bar{J}S+\bar{B}/2) -\bar{\mu} ]/{\bar{T}}}+1}   
                                                  +  \bar{\omega } -y^2+ (\bar{J}S+\frac{\bar{B}}{2}+\bar {\mu})    -  \frac{i}{2\bar{\tau}}     \Bigg]   \nonumber       \\ 
      & - &    \frac{1}{{\bar{\tau}}^2}    \frac{1}{\Big[  \bar{\omega }+ \bar{\Omega } -y^2- (\bar{J}S+\frac{\bar{B}}{2}+\bar {\mu})  \Big]^2 + (\frac{1}{2\bar{\tau}})^2 } 
                                  \cdot     \frac{1}{{\rm{e}}^{[{{y}}^2 +(\bar{J}S+\bar{B}/2) -\bar{\mu} ]/{\bar{T}}}+1}   \nonumber        \\
    &  \times &       \frac{1}{\Big[  \bar{\omega } -y^2+ (\bar{J}S+\frac{\bar{B}}{2}+\bar {\mu})  \Big]^2 + (\frac{1}{2\bar{\tau}})^2 }   
                                     \cdot   \Big(   1-  \frac{1}{{\rm{e}}^{[{{y}}^2 -(\bar{J}S+\bar{B}/2) -\bar{\mu} ]/{\bar{T}}}+1 }  \Big)          
      \Bigg\},       \nonumber         \\
\label{eqn:e100}
\end{eqnarray}
and
\begin{eqnarray}
({\rm{II}})&:=&   \bigg{\langle}    -iJa_0^3  \sqrt{\frac{\tilde S}{2}}    \Big[1-\frac{ a^{\dagger }({\mathbf{x}},t) a({\mathbf{x}},t)}{4 \tilde S }\Big] a({\mathbf{x}},t) 
                                                      c^{\dagger }  ({\mathbf{x}},t) \sigma^{-}   c({\mathbf{x}},t )  \bigg{\rangle} 
   +    \bigg{\langle}  - \frac{\Gamma (t)  c^{\dagger }  ({\mathbf{x}},t) \sigma^{-}   c({\mathbf{x}},t )  }{4i}  \bigg{\rangle}   \nonumber      \\
&=&- \frac{ JS }{ 4 \pi }(\frac{\Gamma _0}{2})^2  \int \frac{d{ {\mathbf{k}}}}{(2\pi )^3}   \int d\omega 
                                  \Big[1-\frac{i}{\tilde S }   \int \frac{d{ {\mathbf{k}}}^{\prime}}{(2\pi )^3}     \int \frac{d{\omega}^{\prime} }{2\pi} 
                                              {\rm{G}}^{\rm{<}}_{{ {\mathbf{k}}}^{\prime}, \omega ^{\prime}}    \Big]         \\ 
& \times & \Bigg{\{}  
    \Big[   {\rm{e}}^{2i\Omega t} {\rm{G}}^{\rm{r}}_{0, \Omega } + {\rm{G}}^{\rm{r}}_{0, -\Omega }    
               + ({\rm{e}}^{2i\Omega t}+1) {\rm{G}}^{\rm{a}}_{0, -\Omega }   
    \Big]  
    \cdot \Big[{\mathcal{G}}^{\rm{t}}_{\uparrow , {\mathbf{k}}, \omega -\Omega } 
              {\mathcal{G}}^{\rm{t}}_{\downarrow  , {\mathbf{k}}, \omega  }                                                        
                -  {\mathcal{G}}^{\rm{<}}_{\uparrow , {\mathbf{k}}, \omega -\Omega } 
              {\mathcal{G}}^{\rm{>}}_{\downarrow  , {\mathbf{k}}, \omega  } 
    \Big]                                                                                                                                                                                 \nonumber       \\
  &+& \Big[   {\rm{e}}^{-2i\Omega t} {\rm{G}}^{\rm{r}}_{0, -\Omega } + {\rm{G}}^{\rm{r}}_{0, \Omega }    
               + ({\rm{e}}^{-2i\Omega t}+1) {\rm{G}}^{\rm{a}}_{0, \Omega }   
    \Big]   
    \cdot  \Big[{\mathcal{G}}^{\rm{t}}_{\uparrow , {\mathbf{k}}, \omega +\Omega } 
     {\mathcal{G}}^{\rm{t}}_{\downarrow  , {\mathbf{k}}, \omega  }   
                 -  {\mathcal{G}}^{\rm{<}}_{\uparrow , {\mathbf{k}}, \omega +\Omega } 
              {\mathcal{G}}^{\rm{>}}_{\downarrow  , {\mathbf{k}}, \omega  } 
    \Big]                                           \Bigg{\}}   \nonumber        \\
    &+& {\cal{O}}(J^2)      + {\cal{O}}({(\Gamma_0)}^0)       \\
  &=&  -    {\bar{J}} \int  d\bar {\omega }  \int_0^{\infty } dy y^2  
        \Bigg[
                   \frac{  {\rm{e}}^{2i\Omega t} (\bar {\omega  }  -\bar{B}-\frac{i}{2\bar{\tau}_{\rm{m}}})  }{  (\bar {\Omega  }  -\bar{B})^2+ (\frac{1}{-2\bar{\tau}_{\rm{m}}})^2  }
                   -  \frac{  ({\rm{e}}^{2i\Omega t} +2) (\bar {\Omega  }  +\bar{B}) - {\rm{e}}^{2i\Omega t}  
                      \frac{i}{2\bar{\tau}_{\rm{m}}})  }{  (\bar {\Omega  }  +\bar{B})^2+ (\frac{1}{-2\bar{\tau}_{\rm{m}}})^2  }         
        \Bigg]    \nonumber        \\
&\times & \Big[1-\frac{i}{\tilde S }  \int \frac{d{ {\mathbf{k}}}^{\prime}}{(2\pi )^3}      \int \frac{d{\omega}^{\prime} }{2\pi} 
                           f_{\rm{B}} (\omega^{\prime})   \frac{-i/{\tau_{\rm{m}}}}{(\omega ^{\prime}-\omega _{{\mathbf{k^{\prime}}}})^2+[1/(2\tau_{\rm{m}})]^2}       
 \Big]   \nonumber    \\ 
     & \times &  \Bigg\{ \frac{1}{\Big[  \bar{\omega }- \bar{\Omega } -y^2+ (\bar{J}S+\frac{\bar{B}}{2}+\bar {\mu})  \Big]^2 + (\frac{1}{2\bar{\tau}})^2 }  \nonumber     \\
    &\times &                                    \Bigg[
                                     \frac{i}{\bar{\tau}}  \frac{1}{{\rm{e}}^{[{{y}}^2 -(\bar{J}S+\bar{B}/2) -\bar{\mu} ]/{\bar{T}}}+1}   
                                                  +  \bar{\omega }- \bar{\Omega } -y^2+ (\bar{J}S+\frac{\bar{B}}{2}+\bar {\mu})    -  \frac{i}{2\bar{\tau}}
                                    \Bigg] \nonumber     \\
        &   \times  & \frac{1}{\Big[  \bar{\omega } -y^2- (\bar{J}S+\frac{\bar{B}}{2}+\bar {\mu})  \Big]^2 + (\frac{1}{2\bar{\tau}})^2 } 
                                    \Bigg[
                                     \frac{i}{\bar{\tau}}  \frac{1}{{\rm{e}}^{[{{y}}^2 +(\bar{J}S+\bar{B}/2) -\bar{\mu} ]/{\bar{T}}}+1}   
                                                  +  \bar{\omega } -y^2- (\bar{J}S+\frac{\bar{B}}{2}+\bar {\mu})    -  \frac{i}{2\bar{\tau}}     \Bigg]   \nonumber      \\ 
     &  - &    \frac{1}{{\bar{\tau}}^2}    \frac{1}{\Big[  \bar{\omega }- \bar{\Omega } -y^2+ (\bar{J}S+\frac{\bar{B}}{2}+\bar {\mu})  \Big]^2 + (\frac{1}{2\bar{\tau}})^2 } 
                                  \cdot     \frac{1}{{\rm{e}}^{[{{y}}^2 -(\bar{J}S+\bar{B}/2) -\bar{\mu} ]/{\bar{T}}}+1}   \nonumber          \\
    &   \times   &     \frac{1}{\Big[  \bar{\omega } -y^2- (\bar{J}S+\frac{\bar{B}}{2}+\bar {\mu})  \Big]^2 + (\frac{1}{2\bar{\tau}})^2 }   
                                     \cdot   \Big(   1-  \frac{1}{{\rm{e}}^{[{{y}}^2 +(\bar{J}S+\bar{B}/2) -\bar{\mu} ]/{\bar{T}}}+1 }  \Big)          
      \Bigg\}  \nonumber        \\
& -  &  {\bar{J}} \int  d\bar {\omega }  \int_0^{\infty } dy y^2  
        \Bigg[
                   \frac{  {\rm{e}}^{-2i\Omega t} (\bar {\omega  }  -\bar{B}-\frac{i}{2\bar{\tau}_{\rm{m}}})  }{  (-\bar {\Omega  }  -\bar{B})^2+ (\frac{1}{-2\bar{\tau}_{\rm{m}}})^2  }
                   -  \frac{  ({\rm{e}}^{-2i\Omega t} +2) (-\bar {\Omega  }  +\bar{B}) - {\rm{e}}^{-2i\Omega t}  
                      \frac{i}{2\bar{\tau}_{\rm{m}}})  }{  (-\bar {\Omega  }  +\bar{B})^2+ (\frac{1}{-2\bar{\tau}_{\rm{m}}})^2  }
        \Bigg]        \nonumber         \\
&\times &    \Big[1-\frac{i}{\tilde S }    \int \frac{d{ {\mathbf{k}}}^{\prime}}{(2\pi )^3}     \int \frac{d{\omega}^{\prime} }{2\pi} 
                    f_{\rm{B}} (\omega^{\prime})   \frac{-i/{\tau_{\rm{m}}}}{(\omega ^{\prime}-\omega _{{\mathbf{k^{\prime}}}})^2+[1/(2\tau_{\rm{m}})]^2}             
 \Big]  \nonumber         \\ 
    &  \times &  \Bigg\{ \frac{1}{\Big[  \bar{\omega }+ \bar{\Omega } -y^2+ (\bar{J}S+\frac{\bar{B}}{2}+\bar {\mu})  \Big]^2 + (\frac{1}{2\bar{\tau}})^2 } \nonumber    \\
      &\times &                               \Bigg[
                                     \frac{i}{\bar{\tau}}  \frac{1}{{\rm{e}}^{[{{y}}^2 -(\bar{J}S+\bar{B}/2) -\bar{\mu} ]/{\bar{T}}}+1}   
                                                  +  \bar{\omega }+ \bar{\Omega } -y^2+ (\bar{J}S+\frac{\bar{B}}{2}+\bar {\mu})    -  \frac{i}{2\bar{\tau}}
                                    \Bigg] \nonumber     \\
         &  \times  & \frac{1}{\Big[  \bar{\omega } -y^2- (\bar{J}S+\frac{\bar{B}}{2}+\bar {\mu})  \Big]^2 + (\frac{1}{2\bar{\tau}})^2 } \nonumber      \\
 &\times &                                    \Bigg[
                                     \frac{i}{\bar{\tau}}  \frac{1}{{\rm{e}}^{[{{y}}^2 +(\bar{J}S+\bar{B}/2) -\bar{\mu} ]/{\bar{T}}}+1}   
                                                  +  \bar{\omega } -y^2- (\bar{J}S+\frac{\bar{B}}{2}+\bar {\mu})    -  \frac{i}{2\bar{\tau}}     \Bigg] \nonumber        \\ 
      & - &    \frac{1}{{\bar{\tau}}^2}    \frac{1}{\Big[  \bar{\omega }+ \bar{\Omega } -y^2+ (\bar{J}S+\frac{\bar{B}}{2}+\bar {\mu})  \Big]^2 + (\frac{1}{2\bar{\tau}})^2 } 
                                  \cdot     \frac{1}{{\rm{e}}^{[{{y}}^2 -(\bar{J}S+\bar{B}/2) -\bar{\mu} ]/{\bar{T}}}+1}   \nonumber          \\
  &     \times   &     \frac{1}{\Big[  \bar{\omega } -y^2- (\bar{J}S+\frac{\bar{B}}{2}+\bar {\mu})  \Big]^2 + (\frac{1}{2\bar{\tau}})^2 }   
                                     \cdot   \Big(   1-  \frac{1}{{\rm{e}}^{[{{y}}^2 +(\bar{J}S+\bar{B}/2) -\bar{\mu} ]/{\bar{T}}}+1 }  \Big)           
      \Bigg\}.   
\label{eqn:e200}      
\end{eqnarray}




\begin{thebibliography}{10}
\bibitem{mod}
I. Zutic, J. Fabian, and S. D. Sarma:
 { Rev. Mod. Phys.}
$\mathbf{76}$
(2004)
323.
\bibitem{mod2}
Y. Tserkovnyak, A. Brataas, G. E. W. Bauer, and B. I. Halperin:
 {Rev. Mod. Phys.}
$\mathbf{77}$
(2005)
1375,
and references therein.
 \bibitem{spinwave}
Y. Kajiwara, K. Harii, S. Takahashi, J. Ohe, K. Uchida,
	M. Mizuguchi, H. Umezawa, H. Kawai, K. Ando, K. Takanashi,
	 S. Maekawa, and E. Saitoh:
 {Nature}
$\mathbf{464}$
(2010)
262.
\bibitem{sandweg}
C. W. Sandweg, Y. Kajiwara, A. V. Chumak, A. A. Serga, V. I. Vasyuchka,
	 M. B. Jungfleisch, E. Saitoh, and B. Hillebrands:
 {Phys. Rev. Lett.}
$\mathbf{106}$
(2011)
216601.
\bibitem{Kurebayashi}
H. Kurebayashi, O. Dzyapko, V. E. Demidov, D. Fang, A. J. Ferguson,
	and S. O. Demokritov:
 {Nat. Mater.}
$\mathbf{10}$
(2011)
660.
\bibitem{interface}
E. Simanek  and B. Heinrich:
 {Phys. Rev. B}
$\mathbf{67}$
(2003)
144418.
\bibitem{tsutsui}
K. Tsutsui, A. Takeuchi, G. Tatara, and S. Murakami:
 {J. Phys. Soc. Jpn.}
$\mathbf{80}$
(2011)
084701.
\bibitem{gene}
S. Zhang and Z. Li:
 {Phys. Rev. Lett.}
$\mathbf{93}$
(2004)
127204.
\bibitem{def}
J. Shi, P. Zhang, X. Xiao, and Q. Niu:
 {Phys. Rev. Lett.}
$\mathbf{96}$
(2006)
076604.
 \bibitem{mista}
L. Mista:
 {Phys. letters}
$\mathbf{25A}$
(1967)
646.
\bibitem{torqueJMM}
R. C. Ralph and M. D. Stiles:
 {J. Magn. Magn. Mater.}
$\mathbf{320}$
(2008)
1190.
 \bibitem{nakatatatara}
K. Nakata and G. Tatara:
 {J. Phys. Soc. Jpn.}
$\mathbf{80}$
(2011)
054602.
 \bibitem{tataraprivate}
G. Tatara (private communication).
\bibitem{AndoPumping}
K. Ando, S. Takahashi, J. Ieda, H. Kurebayashi, T. Trypiniotis,
	 C. H. W. Barnes, S. Maekawa, and E. Saitoh:
 {Nat. Mater.}
$\mathbf{10}$
(2011)
655.
\bibitem{ramer}
J. Rammer and H. Smith:
 {Rev. Mod. Phys.}
 $\mathbf{58}$
(1986)
323.
\bibitem{kamenev}
A. Kamenev:
\textit{Field Theory of Non-Equilibrium Systems}
 (Cambridge University Press, 2011,  arXiv:0412296)
 p. 31.
 \bibitem{kita}
T. Kita:
 {Prog. Theor. Phys.}
$\mathbf{123}$
(2010)
581.
\bibitem{haug}
H. Haug and A. P. Jauho:
\textit{ Quantum Kinetics in Transport and Optics of Semiconductors}
 (Springer New York, 2007)
 p. 66.
\bibitem{tatara}
G. Tatara, H. Kohno,  and J. Shibata:
 { Phys. Rep.}
$\mathbf{468}$ 
(2008)
213.
\bibitem{new}
D. A. Ryndyk, R. Gutierrez, B. Song, and G. Cuniberti:
\textit{Energy Transfer Dynamics in Biomaterial Systems }
 (Springer-Verlag, 2009, arXiv:0805.0628)
 p. 213.
\bibitem{rammer}
J. Rammer:
\textit{Quantum Field Theory of Non-equilibrium States}
 (Cambridge University Press, 2007)
 p. 93.
 \bibitem{adachiseebeck}
H. Adachi, J. Ohe, S. Takahashi, and S. Maekawa:
 {Phys. Rev. B}
$\mathbf{83}$
(2011)
094410.
 \bibitem{takeuchi}
A. Takeuchi, K. Hosono, and G. Tatara:
 {Phys. Rev. B}
$\mathbf{81}$
(2010)
144405.
 \bibitem{xiao}
J. Xiao, G. E. W. Bauer, K. Uchida, E. Saitoh, and S. Maekawa:
 {Phys. Rev. B}
$\mathbf{81}$
(2010)
214418,
and references therein.
 \bibitem{kittel}
C. Kittel:
\textit{Introduction to Solid State Physics}
 (Wiley, 1963)
p. 49.
\bibitem{nakatanote}
K. Nakata:
 {Mod. Phys. Lett. B}
$\mathbf{26}$
(2012)
1250093/arXiv:1204.2339.
\bibitem{Kurebayashi}
H. Kurebayashi, O. Dzyapko, V. E. Demidov, D. Fang, A. J. Ferguson,
	and S. O. Demokritov:
 {Nat. Mater.}
$\mathbf{10}$
(2011)
660.
\bibitem{beta1}
S. Zhang and Z. Li:
 {Phys. Rev. Lett.}
$\mathbf{93}$
(2004)
127204.
\bibitem{beta2}
A. Thiaville, Y. Nakatani, J. Miltat, and Y. Suzuki:
 {Europhys. Lett.}
$\mathbf{69}$
(2005)
990.
\bibitem{pumping5}
K. Ando, Y. Kajiwara, S. Takahashi, S. Maekawa, K. Takemoto, M. Takatsu, and E. Saitoh:
 {Phys. Rev. B}
$\mathbf{78}$
(2008)
014413.
\bibitem{pumping6}
H. Y. Inoue, K. Harii, K. Ando, K. Sasage, and E. Saitoh:
 {J. Appl. Phys.}
$\mathbf{102}$
(2007)
083915.
\bibitem{bauerrev}
A. Brataas, Y. Tserkovnyak, G. E. W. Bauer, and P. J. Kelly:
 {arXiv:1108.0385}.
\bibitem{yunoki}
Q. Zhang, S. Hikino, and S. Yunoki:
 {Appl. Phys. Lett.}
$\mathbf{99}$
(2011)
172105.
\bibitem{conductance}
K. Xia, P. J. Kelly, G. E. W. Bauer, A. Brataas, and I. Turek:
 {Phys. Rev. B}
$\mathbf{65}$
(2002)
220401(R).
\bibitem{Gilbert}
T. L. Gilbert:
 {IEEE Transactions on Magnetics}
$\mathbf{40}$
(2004)
3443.
\bibitem{tatara2}
H. Kohno, G. Tatara, and J. Shibata:
 {J. Phys. Soc. Jpn.}
$\mathbf{75}$
(2006)
113706.
\bibitem{battery}
A. Brataas, Y. Tserkovnyak, G. E. W. Bauer and B. I. Halperin:
 {Phys. Rev. B}
$\mathbf{66}$
(2002)
060404(R).
\end{thebibliography}
\end{document}